 \documentclass[final,1p,times]{elsarticle}

\usepackage{amssymb}
\usepackage{amsthm}
\usepackage{bm}
\usepackage{amsmath}
\usepackage{autobreak}
\usepackage{color}
\usepackage{subfigure}
\usepackage{booktabs}
\usepackage{hyperref}
\biboptions{sort&compress}

\begin{document}

\begin{frontmatter}



\title{A lattice Boltzmann model for the coupled cross-diffusion-fluid system}

\author[a]{Chengjie Zhan}
\author[a,b]{Zhenhua Chai}
\author[a,b]{Baochang Shi \corref{cor1}}
	\ead{shibc@hust.edu.cn}
\address[a]{School of Mathematics and Statistics, Huazhong University of Science and Technology, Wuhan, 430074, China}
\address[b]{Hubei Key Laboratory of Engineering Modeling and Scientific Computing, Huazhong University of Science and Technology, Wuhan 430074, China}
\cortext[cor1]{Corresponding author.}
\begin{abstract}
In this paper, we propose a lattice Boltzmann (LB) model for the generalized coupled cross-diffusion-fluid system. Through the direct Taylor expansion method, the proposed LB model can correctly recover the macroscopic equations. The cross diffusion terms in the coupled system are modeled by introducing additional collision operators, which can be used to avoid special treatments for the gradient terms. In addition, the auxiliary source terms are constructed properly such that the numerical diffusion caused by the convection can be eliminated. We adopt the developed LB model to study two important systems, i.e., the coupled chemotaxis-fluid system and the double-diffusive convection system with Soret and Dufour effects. We first test the present LB model through considering a steady-state case of coupled chemotaxis-fluid system, then we analyze the influences of some physical parameters on the formation of sinking plumes. Finally, the double-diffusive natural convection system with Soret and Dufour effects is also studied, and the numerical results agree well with some previous works. 
\end{abstract}




\end{frontmatter}


\section{Introduction}
\label{Tntro}
Cross-diffusion systems have been widely used to describe multi-species interaction in many fields, for instance, chemotactic cell migration\cite{KS1970JTB,KS1971JTB}, population dynamic in biological systems\cite{Shige1979JTB}, pedestrian dynamics\cite{Hitt2017MMMAS},  multicomponent diffusion \cite{Curtiss1999IECR} and so on. Such systems can also induce some interesting and multiple phenomena, which are significant in both science and engineering. However, most of these systems usually couple with the fluid field, and one also needs to consider the effect of fluid flow. In this work, we consider a general cross-diffusion-fluid (CDF) system in $d$ dimensional space, 
\begin{subequations}\label{CDF}
	\begin{equation}\label{CDE}
	\frac{\partial\phi_{\alpha}}{\partial t}+\nabla\cdot\phi_{\alpha}\mathbf{u}=\nabla\cdot D_{\alpha\beta}\nabla\phi_{\beta}+S_{\alpha},
	\end{equation}
	\begin{equation}\label{NS1}
	\nabla\cdot\mathbf{u}=0,
	\end{equation}
	\begin{equation}\label{NS2}
	\frac{\partial \mathbf{u}}{\partial t} + \nabla\cdot\mathbf{uu}=-\nabla p + \nabla\cdot\nu\nabla\mathbf{u}+\mathbf{F},
	\end{equation}
\end{subequations}
where $\bm{\Phi}=(\phi_{\alpha})$ and $\mathbf{S}=(S_{\alpha})$ are $m\times 1$ matrices, $m$ is the number of scalar variables. $\mathbf{D}=(D_{\alpha\beta})$ is a $m\times m$ diffusive matrix, and $D_{\alpha\beta}\nabla\phi_{\beta}$ means $\sum_{\beta=1}^{m}D_{\alpha\beta}\nabla\phi_{\beta}$ with $\alpha=1, 2, \cdots, m$. $\mathbf{u}$ and $\mathbf{F}$ are the $d$ dimensional fluid velocity and external force, respectively. $p$ is the pressure and $\nu$ is the viscosity. We note that there are two important cases of such a general system, i.e., the coupled chemotaxis-fluid (CF) system \cite{Hillesdon1995BMB,Dombrowski2004PRL,Tuval2005PNAS} and the double-diffusive convection (DDC) system with Soret and Dufour effects \cite{Trevisan1987JHT,Gaikwad2007IJNLM}, which can be obtained through taking the specific forms of $\mathbf{\Phi}$, $\mathbf{D}$ and $\mathbf{F}$.

The CDF system is nonlinear and coupled, and usually it is difficult to obtain its analytical solution. For this reason, some numerical methods have been developed for this type of system \cite{Budroni2015PRE,Bend2018MMAS,KIM2019IJHMT,Rag2019ZAMP,Atlas2020AMM}. For CF system, 
Chertock et al. \cite{Chertock2012JFM} developed a high-resolution vorticity-based hybrid finite-volume finite-difference scheme to understand the interplay of gravity and chemotaxis in the formation of two-dimensional plumes. 
Sheu and Chiang \cite{Sheu2014CF} proposed a combined compact difference scheme of fifth-order spatial accuracy in a three-point grid stencil to investigate the flow convection and diffusion effects on the distributions of the bacteria and oxygen concentrations. 
To avoid a strong restriction on the time step, Lee and Kim \cite{Lee2015EJMBF} used an operator splitting-type Navier-Stokes solver to study the nonlinear dynamics of a three-dimensional CF system. 
Recently, an upwind finite element method was developed to investigate the pattern formation and hydrodynamical stability of the CF system, and through a comparison of chemotaxis-diffusion, double diffusive, and Rayleigh-B\'{e}nard convection, some similarities among them were obtained in \cite{Deleuze2016CF}.
For DDC system with Soret and Dufour effects, 
Nithyadevi and Yang \cite{Nithyadevi2009IJHFF}, using SIMPLE algorithm with QUICK scheme, studied the DDC of water with Soret and Dufour factors in a partially heated enclosure around the density maximum. 
B\'{e}g et al. \cite{Beg2011IJHMT} analyzed the steady free convection heat and mass transfer from a spherical body in a micropolar fluid with Soret and Dufour effects by the Keller-box implicit finite difference method. 
The free convection boundary layer flow over an arbitrarily inclined heated plate in a porous medium with Soret and Dufour effects was studied by transforming the governing equations into a universal form in \cite{Cheng2012ICHMT}.
Wang et al. \cite{Wang2014IJHMT,Wang2016IJTS} studied the oscillation of double-diffusive B\'{e}nard convection with Soret and Dufour effects in a horizontal cavity by the SIMPLE algorithm.
In this work, we will consider the lattice Boltzmann (LB) method for its advantages in the study of complex physical systems.

The LB method, as a mesoscopic numerical approach, has achieved great success in modeling complex flows \cite{Higuera1989EPL,Benzi1992PR,Qian1995ARCP,Chen1998ARFM,Aidun2010ARFM,Wang2019Capillarity} and nonlinear systems, 
such as reaction-diffusion equation \cite{Dawson1993JCP,Blaak2000CPC}, convection-diffusion equation \cite{He2000MS,Shi2009PRE,Chai2013PRE} and so on. In addition, the LB method can describe the coupling interaction among different physical fields well for its mesoscopic kinetic background \cite{Succi2001,Kruger2017}, and it is also suitable to study coupled problems like the present CDF system. 
However, there is a very difficult issue in the study of the CDF system in the framework of the LB method, i.e., how to treat the cross-diffusion terms? Hilpert \cite{Hilpert2005JMB} presented a strategy to put the cross-diffusion term into the equilibrium distribution function, but the finite-difference scheme is needed to compute the gradient operator. Yu et al. \cite{Yu2007ICCS} handled the cross-diffusion terms through redesigning the second-moments of equilibrium distribution functions, thus no gradient operators included. The cross-diffusion term was put into the evolution equation as a source term such that the gradient operator can be computed by a local computational scheme in Refs. \cite{Yang2014CMA,Chai2019SIAM}. Ren and Chan \cite{Ren2016IJHMT} diagonalized the coupling diffusivities matrix, thus the coupled system can be transformed to the uncoupled convection-diffusion equations.

In this work, we will develop a LB model for the CDF system.
Inspired by Refs.\cite{Huber2010JCP,Chai2019PRE} where the source terms and fluid field have not been included, to avoid the special treatments for the gradient terms, as mentioned above, the cross-diffusion terms in the coupled system are modeled by some extra collision operators, which are added in the evolution of LB model. 

This paper is organized as follows. In Section \ref{LBM}, the coupled LB model for general CDF system is proposed. Through the direct Taylor expansion, the governing equations can be correctly recovered from present LB model. In Section \ref{Numerical}, we simulate the coupled CF system and the DDC system with Soret and Dufour effects. For the former system, after testing the LB model, we analyze the influences of some parameters on the formation of plume structures. We also consider the effects of some physical parameters for the latter problem. Finally, some conclusions are summarized in Section \ref{Conclusions}.

\section{The coupled lattice Boltzmann model}
\label{LBM}
In this section we will develop a LB model for the CDF system (\ref{CDF}), and write the evolution equations of present LB model as
\begin{subequations}\label{evolution}
	\begin{align}\label{evolution12}
		\begin{autobreak}
		f_{i,\alpha}\left(\mathbf{x}+\mathbf{c}_{i} \Delta t, t+\Delta t\right)=
		f_{i,\alpha}(\mathbf{x}, t)
		-\omega_{\alpha\beta}\left(f_{i,\beta}(\mathbf{x}, t)-f_{i,\beta}^{eq}(\mathbf{x}, t)\right) 
		+\Delta t S_{i,\alpha}(\mathbf{x}, t)
		+\Delta t G_{i,\alpha}(\mathbf{x}, t)
		+\frac{\Delta t^{2}}{2} \bar{D}_{i,\alpha\beta} S_{i,\beta}(\mathbf{x}, t),
		\end{autobreak}
	\end{align}
	\begin{equation}\label{evolution3}
		h_{i}\left(\mathbf{x}+\mathbf{c}_{i} \Delta t, t+\Delta t\right) =
		h_{i}(\mathbf{x}, t)-\omega\left(h_{i}(\mathbf{x}, t)-h_{i}^{e q}(\mathbf{x}, t)\right)+\Delta t (1-\frac{\omega}{2}) F_{i}(\mathbf{x}, t),
	\end{equation}
\end{subequations}
where $f_{i,\alpha}(\mathbf{x}, t)$ ($i=0,1,\cdots, q-1$, $q$ represents the number of discrete velocity directions) and $h_{i}(\mathbf{x}, t)$ are the distribution functions of scalar variable $\phi_{\alpha}$ and fluid field at position $\mathbf{x}$ and time $t$, $f_{i\alpha}^{eq}(\mathbf{x}, t)$ and $h_{i}^{eq}(\mathbf{x}, t)$ are the corresponding equilibrium distribution functions. $\mathbf{c}_{i}$ is the discrete velocity, $\Delta t$ is the time step. $(\omega_{\alpha\beta})$ is a invertible $m\times m$ matrix, $\omega_{\alpha\beta}$ and $\omega$ are relaxation factors. $\bar{D}_{i,\alpha\beta}=\delta_{\alpha\beta} \partial_{t}+\gamma_{\alpha\beta}\mathbf{c}_{i}\cdot\nabla$ with $(\delta_{\alpha\beta})$ and $(\gamma_{\alpha\beta})$ representing $m\times m$ identity and diagonal matrices, respectively.   

To obtain the macroscopic equation (\ref{CDF}), the equilibrium distribution functions are defined by
\begin{subequations}\label{feq}
	\begin{equation}\label{feq-phi}
	f_{i,\alpha}^{eq}=W_{i}\left[\phi_{\alpha}+\frac{\mathbf{c}_{i}\cdot\phi_{\alpha}\mathbf{u}}{c_{s}^{2}}+\frac{[(\lambda_{\alpha\beta}\phi_{\beta}-\phi_{\alpha})c_{s}^{2}\mathbf{I}+\vartheta\phi_{\alpha}\mathbf{uu}]:(\mathbf{c}_{i}\mathbf{c}_{i}-c_{s}^{2}\mathbf{I})}{2c_{s}^{4}}\right],
	\end{equation}
	\begin{equation}\label{heq}
	h_{i}^{eq}=\sigma_{i}+W_{i}\left[\frac{\mathbf{c}_{i}\cdot \mathbf{u}}{c_{s}^{2}}+\frac{\mathbf{uu}:(\mathbf{c}_{i}\mathbf{c}_{i}-c_{s}^{2}\mathbf{I})}{2c_{s}^{4}}\right],
	\end{equation}
\end{subequations}
where $W_{i}$ is the weight coefficient, and $c_{s}$ is the lattice sound speed in LB method. $(\lambda_{\alpha\beta})$ is a $m\times m$ invertible matrix that can be used to adjust the relaxation matrix $(\omega_{\alpha\beta})$, $\vartheta$ is an adjustable parameter. $\sigma_{0}=(W_{0}-1)p/c_{s}^{2}+\rho_{0}$ with $\rho_{0}$ being a constant, $\sigma_{i}=W_{i}p/c_{s}^{2}\,(i\neq 0)$ \cite{He2004CP}. 

The force term $F_{i}(\mathbf{x}, t)$ is given by
\begin{equation}\label{force}
F_{i}=W_{i}\left[\frac{\mathbf{c}_{i}\cdot\mathbf{F}}{c_{s}^{2}}+\varphi\frac{(\mathbf{uF}+\mathbf{Fu}):(\mathbf{c}_{i}\mathbf{c}_{i}-c_{s}^{2}\mathbf{I})}{2c_{s}^{4}}\right],
\end{equation} 
where $\varphi$ is another adjustable parameter. Here the source term $S_{i,\alpha}(\mathbf{x}, t)$ and auxiliary source term $G_{i,\alpha}(\mathbf{x}, t)$ to be determined later.

The macroscopic scalar variable, fluid velocity and pressure can be computed by
\begin{subequations}\label{sum}
	\begin{equation}\label{sum-phi}
	\phi_{\alpha}=\sum_{i}f_{i,\alpha},
	\end{equation}
	\begin{equation}\label{sum-u}
	\mathbf{u}=\sum_{i}\mathbf{c}_{i}h_{i}+\frac{\Delta t}{2}\mathbf{F},
	\end{equation}
	\begin{equation}\label{sum-p}
	p=\frac{c_{s}^{2}}{1-W_{0}}\left[\sum_{i\neq 0}h_{i}-W_{0}\frac{\mathbf{u}\cdot\mathbf{u}}{2c_{s}^{2}}+\varphi\Delta t(\frac{1}{\omega}-\frac{1}{2})\frac{\mathbf{u}\cdot\mathbf{F}}{c_{s}^{2}}\right],
	\end{equation}
\end{subequations} 
where the details on the computation of pressure can be found in \ref{appendix}.

We noted that the present LB model can be used for $d$ dimensional problems. Here some commonly used lattice velocity models are listed below:
	
D1Q3:
\begin{subequations}\label{D1Q3}
	\begin{equation}
	\mathbf{c}=\hat{c}[0,1,-1],
	\end{equation}
	\begin{equation}
	c_{s}^{2}=\hat{c}^{2}/3,\quad W_{0}=2/3,\quad W_{1-2}=1/6,
	\end{equation}	
\end{subequations}

D2Q5:
\begin{subequations}\label{D2Q5}
	\begin{equation}
	\mathbf{c}=\hat{c}\begin{bmatrix}
	0 & 1 & 0 & -1 & 0 \\
	0 & 0 & 1 & 0 & -1
	\end{bmatrix},
	\end{equation}
	\begin{equation}
	c_{s}^{2}=\hat{c}^{2}/3,\quad W_{0}=1/3,\quad W_{1-4}=1/6,
	\end{equation}
\end{subequations}

D2Q9:
\begin{subequations}\label{D2Q9}
	\begin{equation}
	\mathbf{c}=\hat{c}\begin{bmatrix}
	0 & 1 & 0 & -1 & 0 & 1 & -1 & -1 & 1\\
	0 & 0 & 1 & 0 & -1 & 1 & 1 & -1 & -1
	\end{bmatrix},
	\end{equation}
	\begin{equation}
	c_{s}^{2}=\hat{c}^{2}/3,\quad W_{0}=4/9,\quad W_{1-4}=1/9,\quad W_{5-8}=1/36,
	\end{equation}
\end{subequations}
	
D3Q7:
\begin{subequations}\label{D3Q7}
	\begin{equation}
	\mathbf{c}=\hat{c}\begin{bmatrix}
	0 & 1 & 0 & 0 & -1 & 0 & 0\\
	0 & 0 & 1 & 0 & 0 & -1 & 0\\
	0 & 0 & 0 & 1 & 0 & 0 & -1
	\end{bmatrix},
	\end{equation}
	\begin{equation}
	c_{s}^{2}=\hat{c}^{2}/4,\quad W_{0}=1/4,\quad W_{1-6}=1/8,
	\end{equation}
\end{subequations}

D3Q15:
\begin{subequations}\label{D3Q15}
	\setcounter{MaxMatrixCols}{15}
	\begin{equation}
	\mathbf{c}=\hat{c}\begin{bmatrix}
	0 & 1 & 0 & 0 & -1 & 0 & 0 & 1 & 1 & 1 & -1 & -1 & -1 & -1 & 1\\
	0 & 0 & 1 & 0 & 0 & -1 & 0 & 1 & 1 & -1 & 1 & -1 & -1 & 1 & -1\\
	0 & 0 & 0 & 1 & 0 & 0 & -1 & 1 & -1 & 1 & 1 & -1 & 1 & -1 & -1
	\end{bmatrix},
	\end{equation}
	\begin{equation}
	c_{s}^{2}=\hat{c}^{2}/3,\quad W_{0}=2/9,\quad W_{1-6}=1/9,\quad W_{7-15}=1/72,
	\end{equation}
\end{subequations}

D3Q19:
\begin{subequations}\label{D3Q19}
	\begin{multline}
	\mathbf{c}=\hat{c}\left[\begin{array}{ccccccc}
	0 & 1 & -1 & 0 & 0 & 0 & 0\\
	0 & 0 & 0 & 1 & -1 & 0 & 0\\
	0 & 0 & 0 & 0 & 0 & 1 & -1 
	\end{array}\right.\\
	\left.\begin{array}{cccccccccccc}
	1 & -1 & 1 & -1 & 1 & -1 & -1 & 1 & 0 & 0 & 0 & 0\\
	1 & -1 & -1 & 1 & 0 & 0 & 0 & 0 & 1 & -1 & 1 & -1\\
	0 & 0 & 0 & 0 & 1 & -1 & 1 & -1 & 1 & -1 & -1 & 1
	\end{array}\right],
	\end{multline}
	\begin{equation}
	c_{s}^{2}=\hat{c}^{2}/3,\quad W_{0}=1/3,\quad W_{1-6}=1/18,\quad W_{7-18}=1/36,
	\end{equation}
\end{subequations}
where $\hat{c}=\Delta x/\Delta t$ is the particle speed, and $\Delta x$ is the lattice spacing. 

\subsection{The direct Taylor expansion of present lattice Boltzmann model}
We now perform a direct Taylor expansion analysis of LB model for convection cross-diffusion equation, while the derivation process for incompressible Navier-Stokes equations is shown in \ref{appendix}. In the direct Taylor expansion \cite{Holdych2004JCP,Wagner2006PRE,Kaehler2013CCP,Chai2020PRE}, the time step $\Delta t$ is used as the small expansion parameter.

Applying the Taylor expansion to Eq.\,(\ref{evolution12}) and based on $f_{i,\alpha}=f_{i,\alpha}^{eq}+f_{i,\alpha}^{ne}$ ($f_{i,\alpha}^{ne}$ is the non-equilibrium part of distribution function $f_{i,\alpha}$), we have \cite{Chai2020PRE}
\begin{equation}\label{taylorEx}
\sum_{l=1}^{N}\frac{\Delta t^{l}}{l!}D_{i}^{l}f_{i,\alpha}+O(\Delta t^{N+1})=-\omega_{\alpha\beta}f_{i,\beta}^{ne}+\Delta t S_{i,\alpha}+\Delta t G_{i,\alpha}+\frac{\Delta t^2}{2}\bar{D}_{i,\alpha\beta}S_{i,\beta},
\end{equation}
where $D_{i}=\partial_{t}+\mathbf{c}_{i} \cdot \nabla$. 
From above equation one can obtain
\begin{subequations}\label{taylorEX1}
	\begin{equation}\label{taylorO1}
	f_{i,\alpha}^{ne}=O(\Delta t),
	\end{equation}
	\begin{align}\label{taylorON}
	\begin{autobreak}
	\sum_{l=1}^{N-1}\frac{\Delta t^{l}}{l!}D_{i}^{l}(f_{i,\alpha}^{eq}+f_{i,\alpha}^{ne})
	+\frac{\Delta t^{N}}{N!}D_{i}^{N}f_{i,\alpha}^{eq}=
	-\omega_{\alpha\beta}f_{i,\beta}^{ne}+\Delta t S_{i,\alpha}
	+\Delta t G_{i,\alpha}
	+\frac{\Delta t^2}{2}\bar{D}_{i,\alpha\beta}S_{i,\beta} 
	+ O(\Delta t^{N+1}).
	\end{autobreak}
	\end{align}
\end{subequations}
According to Eq.\,(\ref{taylorEX1}), we can derive the equations at different orders of $\Delta t$,
\begin{subequations}\label{order}
	\begin{equation}\label{order1dt}
	D_{i}f_{i,\alpha}^{eq}=-\frac{\omega_{\alpha\beta}}{\Delta t}f_{i,\beta}^{ne}+S_{i,\alpha}+G_{i,\alpha}+O(\Delta t),
	\end{equation}
	\begin{equation}\label{order2dt}
	D_{i}(f_{i,\alpha}^{eq}+f_{i,\alpha}^{ne})+\frac{\Delta t}{2}D_{i}^{2}f_{i,\alpha}^{eq}=-\frac{\omega_{\alpha\beta}}{\Delta t}f_{i,\beta}^{ne}+S_{i,\alpha}+G_{i,\alpha}+\frac{\Delta t}{2}\bar{D}_{i,\alpha\beta}S_{i,\beta}+O(\Delta t^{2}).
	\end{equation}
\end{subequations}
From Eq.\,(\ref{order1dt}), we can get
\begin{equation}\label{D-order1}
\frac{\Delta t}{2}D_{i}^{2}f_{i,\alpha}^{eq}=-\frac{1}{2}D_{i}\omega_{\alpha\beta}f_{i,\beta}^{ne}+\frac{\Delta t}{2}D_{i}(S_{i,\alpha}+G_{i,\alpha})+O(\Delta t^2).
\end{equation}
Substituting Eq.\,(\ref{D-order1}) into Eq.\,(\ref{order2dt}) yields
\begin{align}\label{order2dt-1}
\begin{autobreak}
D_{i}f_{i,\alpha}^{eq}+D_{i}(\delta_{\alpha\beta}-\frac{\omega_{\alpha\beta}}{2})f_{i,\beta}^{ne}
+\frac{\Delta t}{2}D_{i}(S_{i,\alpha}+G_{i,\alpha})
=-\frac{\omega_{\alpha\beta}}{\Delta t}f_{i,\beta}^{ne}
+S_{i,\alpha}+G_{i,\alpha}
+\frac{\Delta t}{2}\bar{D}_{i,\alpha\beta}S_{i,\beta}
+O(\Delta t^{2}).
\end{autobreak}
\end{align}
To recover the macroscopic equation (\ref{CDE}), the distribution functions $f_{i,\alpha}$, $f_{i,\alpha}^{eq}$, $S_{i,\alpha}$ and $G_{i,\alpha}$ should satisfy following conditions,
\begin{subequations}\label{moment}
	\begin{equation}
	\sum_{i} f_{i\alpha}=\sum_{i} f_{i\alpha}^{e q}=\phi_{\alpha},\ \sum_{i} \mathbf{c}_{i} f_{i\alpha}^{e q}=\phi_{\alpha}\mathbf{u},\ \sum_{i} \mathbf{c}_{i} \mathbf{c}_{i} f_{i\alpha}^{e q}=\lambda_{\alpha\beta}\phi_{\beta}c_{s}^{2}\mathbf{I} +\vartheta\phi_{\alpha}\mathbf{uu},
	\end{equation}
	\begin{equation}
	\sum_{i} S_{i,\alpha}=S_{\alpha},\quad \sum_{i} \mathbf{c}_{i} S_{i,\alpha}=\mathbf{M}_{1S,\alpha}, 
	\end{equation}
	\begin{equation}  
	\sum_{i} G_{i,\alpha}=0,\quad \sum_{i} \mathbf{c}_{i} G_{i,\alpha}=\mathbf{M}_{1G,\alpha},
	\end{equation}
\end{subequations}
which can also be used to derive the following equation,
\begin{equation}
\sum_{i}f_{i,\alpha}^{ne}=\sum_{i}f_{i,\alpha}-\sum_{i}f_{i,\alpha}^{eq}=0.
\end{equation}
Summing Eqs.\,(\ref{order1dt}) and (\ref{order2dt-1}) over $i$ and using above relations, one can obtain
\begin{subequations}
	\begin{equation}\label{order1}
	\partial_{t} \phi_{\alpha}+\nabla \cdot \phi_{\alpha}\mathbf{u}=S_{\alpha}+O(\Delta t),
	\end{equation}
	\begin{align}\label{order2}
	\begin{autobreak}
	\partial_{t} \phi_{\alpha}+\nabla\cdot\phi_{\alpha}\mathbf{u}
	+\nabla \cdot(\delta_{\alpha\theta}-\frac{\omega_{\alpha\theta}}{2}) \sum_{i} \mathbf{c}_{i} f_{i,\theta}^{ne}
	=S_{\alpha}
	+\frac{\Delta t}{2} \nabla\cdot\left[(\gamma_{\alpha\beta}-\delta_{\alpha\beta})\mathbf{M}_{1S,\beta}-\mathbf{M}_{1G,\alpha}\right]
	+O(\Delta t^2),
	\end{autobreak}
	\end{align}
\end{subequations}
where the term $\sum_{i} \mathbf{c}_{i}f_{i,\theta}^{ne}$ can be derived from Eq.\,(\ref{order1dt}),
\begin{equation}
	\begin{aligned}\label{cifi1}
	\sum_{i} \mathbf{c}_{i}f_{i,\theta}^{ne}&=-\omega_{\theta\beta}^{-1} \Delta t\sum_{i} \mathbf{c}_{i}\left[D_{i}f_{i,\beta}^{e q}- S_{i,\beta}-G_{i,\beta}\right]+O(\Delta t^2) \\ &=-\omega_{\theta\beta}^{-1} \Delta t\left[\partial_{t} \phi_{\beta} \mathbf{u}+\nabla\cdot\vartheta\phi_{\beta}\mathbf{uu}+c_{s}^{2}\lambda_{\beta\kappa} \nabla\phi_{\kappa}-\mathbf{M}_{1S,\beta}-\mathbf{M}_{1G,\beta}\right]+O(\Delta t^2),
	\end{aligned}
\end{equation}
where $(\omega_{\theta\beta}^{-1})$ is defined as the inverse matrix of $(\omega_{\theta\beta})$, i.e., $\sum_{\theta}\omega_{\alpha\theta}\omega_{\theta\beta}^{-1}=\delta_{\alpha\beta}$. 

Substituting Eq.\,(\ref{cifi1}) into Eq.\,(\ref{order2}), we get
\begin{equation}\label{order2-1}
\partial_{t}\phi_{\alpha}+\nabla\cdot\phi_{\alpha}\mathbf{u}=\nabla \cdot(\omega_{\alpha\beta}^{-1}-\frac{\delta_{\alpha\beta}}{2})\lambda_{\beta\theta} c_{s}^{2}\Delta t \nabla \phi_{\theta}+S_{\alpha}+\frac{\Delta t}{2} \nabla \cdot \mathbf{RH}_{\alpha}+O(\Delta t^2),
\end{equation}
where
\begin{equation}\label{RH}
\begin{aligned}
\mathbf{RH}_{\alpha}=(2\omega_{\alpha\beta}^{-1}-\delta_{\alpha\beta})\left(\partial_{t}\phi_{\beta}\mathbf{u}+\nabla\cdot\vartheta\phi_{\beta}\mathbf{uu}\right)+(\gamma_{\alpha\beta}-2\omega_{\alpha\beta}^{-1})\mathbf{M}_{1S,\beta}-2\omega_{\alpha\beta}^{-1}\mathbf{M}_{1G,\beta}.
\end{aligned}
\end{equation}
If $\mathbf{RH}_{\alpha}=\mathbf{0}$ or $\mathbf{RH}_{\alpha}=O(\Delta t)$, we can correctly recover the convection cross-diffusion equation (\ref{CDE}) at the order of $\Delta t^2$,
\begin{equation}\label{order2-2}
\partial_{t}\phi_{\alpha}+\nabla\cdot\phi_{\alpha}\mathbf{u}=\nabla\cdot D_{\alpha\beta}\nabla \phi_{\beta}+S_{\alpha}+O(\Delta t^2),
\end{equation}
where 
\begin{equation}\label{omega-D}
D_{\alpha\beta}=(\omega_{\alpha\theta}^{-1}-\frac{\delta_{\alpha\theta}}{2})\lambda_{\theta\beta} c_{s}^{2} \Delta t.
\end{equation}
When taking $\mathbf{M}_{1S,\alpha}=0$, the term $S_{i,\alpha}$ can be defined by
\begin{equation}\label{source}
S_{i,\alpha}=W_{i}S_{\alpha},
\end{equation}
and $(\gamma_{\alpha\beta})$ can be set as zero matrix for simplicity, thus $\bar{D}_{i,\alpha\beta}=\delta_{\alpha\beta} \partial_{t}$, and a first-order explicit finite-difference scheme can be applied, i.e.,  $\partial_{t}S_{i,\alpha}(\mathbf{x}, t)=[S_{i,\alpha}(\mathbf{x}, t)-S_{i,\alpha}(\mathbf{x}, t-\Delta t)]/\Delta t$. 
In this case, the term $\mathbf{RH}_{\alpha}$ can be written as
\begin{equation}
\mathbf{RH}_{\alpha}=(2\omega_{\alpha\beta}^{-1}-\delta_{\alpha\beta})\left(\partial_{t} \phi_{\beta}\mathbf{u}+\nabla\cdot \vartheta\phi_{\beta}\mathbf{uu}\right)-2\omega_{\alpha\beta}^{-1} \mathbf{M}_{1G,\beta}.
\end{equation}
In the following, two cases are considered according to whether the flow field is coupled or not.

\noindent\textbf{Case 1:} If we only consider the convection cross-diffusion equation (\ref{CDE}) without including the fluid field, one can set $\vartheta=0$ and $\lambda_{\alpha\beta}=\delta_{\alpha\beta}$, thus the equilibrium distribution function $f_{i,\alpha}^{eq}$ is linear, and $DdQ(2d+1)$ velocity models can be used for simplicity. Under condition of $\mathbf{RH}_{\alpha}=0$, we get
\begin{equation}
\mathbf{M}_{1G,\alpha}=(\delta_{\alpha\beta}-\frac{\omega_{\alpha\beta}}{2})\partial_{t}\phi_{\beta}\mathbf{u}.
\end{equation}
In this case the term $G_{i,\alpha}$ can be given by
\begin{equation}
G_{i,\alpha}=(\delta_{\alpha\beta}-\frac{\omega_{\alpha\beta}}{2})\frac{W_{i}\mathbf{c}_{i}\cdot\partial_{t}\phi_{\beta}\mathbf{u}}{c_{s}^{2}},
\end{equation}
where the time derivative can be discretized by a first-order explicit finite-difference scheme \cite{Chai2016JSC}.
 
\noindent\textbf{Case 2:} When the incompressible fluid field is considered, the value of $\vartheta$ can be arbitrary since $\mathbf{uu}$ in equilibrium distribution function $f_{i,\alpha}^{eq}$ is the order of $O(Ma^2)$, and can be neglected, thus $f_{i,\alpha}^{eq}$ can also become linear with a special value $\lambda_{\alpha\beta}=\delta_{\alpha\beta}$. In addition, if we take $\vartheta=1$, and with the help of Eqs.\,(\ref{order1}) and (\ref{order12NS}), $\mathbf{RH}_{\alpha}$ can be written as
\begin{equation}
	\begin{aligned}
	\mathbf{RH}_{\alpha}=&(2\omega_{\alpha\beta}^{-1}-\delta_{\alpha\beta})\left[\mathbf{u}\partial_{t}\phi_{\beta}+\phi_{\beta}\partial_{t}\mathbf{u}+\mathbf{u}\nabla\cdot\phi_{\beta}\mathbf{u}+\phi_{\beta}\nabla\cdot\mathbf{uu}\right]-2\omega_{\alpha\beta}^{-1} \mathbf{M}_{1G,\beta}\\
	=&(2\omega_{\alpha\beta}^{-1}-\delta_{\alpha\beta})\left[\mathbf{u}S_{\beta}+\phi_{\beta}(\mathbf{F}-\nabla p)\right]-2\omega_{\alpha\beta}^{-1} \mathbf{M}_{1G,\beta}+O(\Delta t).
	\end{aligned}
\end{equation}
It is clear that the term $\mathbf{RH}_{\alpha}$ is the order of $\Delta t$, if $\mathbf{M}_{1G,\alpha}$ is taken as
\begin{equation}
\mathbf{M}_{1G,\alpha}=(\delta_{\alpha\beta}-\frac{\omega_{\alpha\beta}}{2})\left[\mathbf{u}S_{\beta}+\phi_{\beta}(\mathbf{F}-\nabla p)\right].
\end{equation}
Under the incompressible condition, $\nabla p$ is also the order of $O(Ma^2)$. Keep this in mind, the term $G_{i,\alpha}$ can also be given by
\begin{equation}\label{Gi}
G_{i,\alpha}=(\delta_{\alpha\beta}-\frac{\omega_{\alpha\beta}}{2})\frac{W_{i}\mathbf{c}_{i}\cdot (\mathbf{u}S_{\beta}+\phi_{\beta}\mathbf{F})}{c_{s}^{2}}.
\end{equation}

From above discussion, one can find that there are no special treatments needed for the cross-diffusion terms in the present LB model, and the discretization of time derivative in auxiliary source term also can be avoided when the fluid field is considered.

\section{Numerical experiments}\label{Numerical}
In this section, we will simulate the CF system and DDC system with Soret and Dufour effects in two-dimensional space. we first test proposed LB model, and then discuss the effects of some physical parameters in this two systems. 
In the simulations below, we set $(\lambda_{\alpha\beta})$ as identity matrix and $\varphi=0$ for simplicity, and adopt the D2Q5 model for the convection cross-diffusion equations and the D2Q9 model for the Navier-Stokes equations. The relaxation factor $\omega_{\alpha\beta}$ is computed by Eq.\,(\ref{omega-D}), the anti-bounce-back scheme \cite{Zhang2012PRE} is used to treat Dirichlet boundary conditions of convection cross-diffusion equations, and the half-way bounce-back scheme \cite{Ladd1994JFM1,Ladd1994JFM2} is applied for the other no-flux and velocity boundary conditions. 
\subsection{Numerical results and discussion on the chemotaxis-fluid system}
From system (\ref{CDF}), we can obtain the coupled CF system through taking $\mathbf{\Phi}=(n,c)^{T}$, $\mathbf{D}=\begin{bmatrix}
D_{n} & -\mu r(c)n\\
0 & D_{c}
\end{bmatrix}$, $\mathbf{S}=(0,-\kappa r(c)n)^{T}$, $\mathbf{F}=\mathbf{g}n V_{b}(\rho_{b}-\rho_{0})/\rho_{0}$ and $p=p/\rho_{0}$,
\begin{subequations}\label{KS-NS}
	\begin{equation}
	n_{t}+\nabla\cdot n\mathbf{u}=\nabla \cdot [D_{n}\nabla n-\mu r(c)n\nabla c],
	\end{equation}
	\begin{equation}
	c_{t}+\nabla\cdot c\mathbf{u}=\nabla \cdot D_{c}\nabla c-\kappa r(c)n,
	\end{equation}
	\begin{equation}
	\nabla \cdot \mathbf{u}=0,
	\end{equation}
	\begin{equation}
	\mathbf{u}_{t} + \mathbf{u} \cdot \nabla \mathbf{u}=-\nabla p/\rho_{0} + \nabla \cdot\nu \nabla \mathbf{u}+\mathbf{g}n V_{b}(\rho_{b}-\rho_{0})/\rho_{0},
	\end{equation}
\end{subequations}
where $n$ and $c$ are the concentrations of bacteria and oxygen, respectively. $\mu$ is the chemotactic sensitivity, $\kappa$ is the consumption rate of oxygen, $D_{n}$ and $D_{c}$ are the diffusion coefficients of bacteria and oxygen.  $\rho_{0}$ is the pure fluid density, $\mathbf{g}$ is the gravitation acceleration, $V_{b}$ and $\rho_{b}$ are the volume and density of bacteria, respectively. $r(c)$ is a dimensionless truncated function, which is relevant to the chemotaxis cut-off value $c^{*}$. 

Consider this system in two-dimensional rectangular domain $\Omega=[-a,a]\times[0,d]$, the boundary conditions are given as follows. The top of the domain $\partial\Omega_{top}$ is fluid-air surface where the flux of bacteria is zero, the value of oxygen concentration equals to the air oxygen concentration $c_{air}$ and no fluid flows through the fluid-air surface,
\begin{equation}
[D_{n}\nabla n-\mu r(c)n\nabla c]\cdot \hat{\mathbf{n}}=0, \ c=c_{air}, \ 
\frac{\partial{u}}{\partial{y}}=0, \ v=0,
\forall\mathbf{x} \in \partial \Omega_{top},
\end{equation}
where $\hat{\mathbf{n}}$ is the unit outer normal vector, $c_{air}$ is the air oxygen concentration, $\mathbf{x}=(x,y)$ and $\mathbf{u}=(u,v)$. At the bottom of the domain $\partial\Omega_{bot}$, the fluxes of bacteria and oxygen, and the fluid velocity are zero,
\begin{equation}
\nabla n\cdot \hat{\mathbf{n}}=\nabla c\cdot \hat{\mathbf{n}}=0, \quad \mathbf{u}=\mathbf{0}, \quad \forall\mathbf{x} \in \partial \Omega_{bot}.
\end{equation}
Periodic boundary condition is used for the left and right sides of the domain.

Before performing any simulations, we first rewrite the system (\ref{KS-NS}) in a dimensionless form through introducing the following variables \cite{Tuval2005PNAS,Chertock2012JFM}:
\begin{equation}
\mathbf{x}'=\frac{\mathbf{x}}{L},\quad t'=\frac{D_{n}}{L^{2}}t,\quad c'=\frac{c}{c_{air}},\quad n'=\frac{n}{n_{r}},\quad \mathbf{u}'=\frac{L}{D_{n}}\mathbf{u},\quad p'=\frac{L^{2}}{\nu\rho_{0} D_{n}}p,
\end{equation}
where $L$ and $n_{r}$ are characteristic length and characteristic bacteria density, respectively. After dropping the prime notation in the rescaled variables, we can obtain the following dimensionless system:
\begin{subequations}\label{non-KS-NS}
	\begin{equation}\label{non-n}
	n_{t}+\nabla\cdot n\mathbf{u}=\nabla \cdot[\nabla n- \alpha r(c)n\nabla c],
	\end{equation}
	\begin{equation}\label{non-c}
	c_{t}+\nabla\cdot c\mathbf{u}=\nabla \cdot\delta\nabla c-\beta r(c)n,
	\end{equation}
	\begin{equation}\label{non-NS11}
	\nabla \cdot \mathbf{u}=0,
	\end{equation}
	\begin{equation}\label{non-NS21}
	\mathbf{u}_{t} + \nabla \cdot \mathbf{uu}=-Sc\nabla p + \nabla \cdot Sc \nabla \mathbf{u}-Sc\gamma n\hat{\mathbf{z}},
	\end{equation}
\end{subequations}
where $\hat{\mathbf{z}}$ is the upwards unit vector, $\alpha$, $\beta$, $\delta$, $\gamma$ and Schmidt number $Sc$ are dimensionless parameters, which are defined as
\begin{equation}
\alpha=\frac{\mu c_{a i r}}{D_{n}}, \ \beta=\frac{\kappa n_{r} L^{2}}{c_{a i r} D_{n}}, \  \delta=\frac{D_{c}}{D_{n}}, \ \gamma=\frac{V_{b} n_{r} g\left(\rho_{b}-\rho_{0}\right) L^{3}}{\nu \rho_{0}D_{n}}, \ Sc=\frac{\nu}{D_{n}}.
\end{equation}

Similarly, we can also derive the dimensionless boundary conditions,
\begin{equation}\label{top-boundary}
[\nabla n-\alpha r(c)n\nabla c]\cdot \hat{\mathbf{n}}=0, \ c=1, \ \mathbf{u}\cdot\hat{\mathbf{n}}=0, \
\forall\mathbf{x} \in \partial \Omega_{top},
\end{equation}
\begin{equation}\label{bot-boundary}
\nabla n\cdot \hat{\mathbf{n}}=\nabla c\cdot \hat{\mathbf{n}}=0, \quad \mathbf{u}=\mathbf{0}, \quad \forall\mathbf{x} \in \partial \Omega_{bot},
\end{equation}
the periodic boundary condition is still used for the left and right boundaries.

Now we can simulate the dimensionless CF system with proposed LB model by setting $\mathbf{\Phi}=(n,c)^{T}$, $\mathbf{D}=\begin{bmatrix}
1 & -\alpha r(c)n\\
0 & \delta
\end{bmatrix}$, $\mathbf{S}=(0,-\beta r(c)n)^{T}$, $\nu=Sc$ and $\mathbf{F}=-Sc\gamma n\hat{\mathbf{z}}$. 

\subsubsection{The validation of present lattice Boltzmann model}\label{verification}
 \begin{figure}
	\setlength{\abovecaptionskip}{-0.05cm} 
	\setlength{\belowcaptionskip}{-0.3cm} 
	\subfigure[]{
		\begin{minipage}{0.5\linewidth}
			\centering
			\includegraphics[width=2.5in]{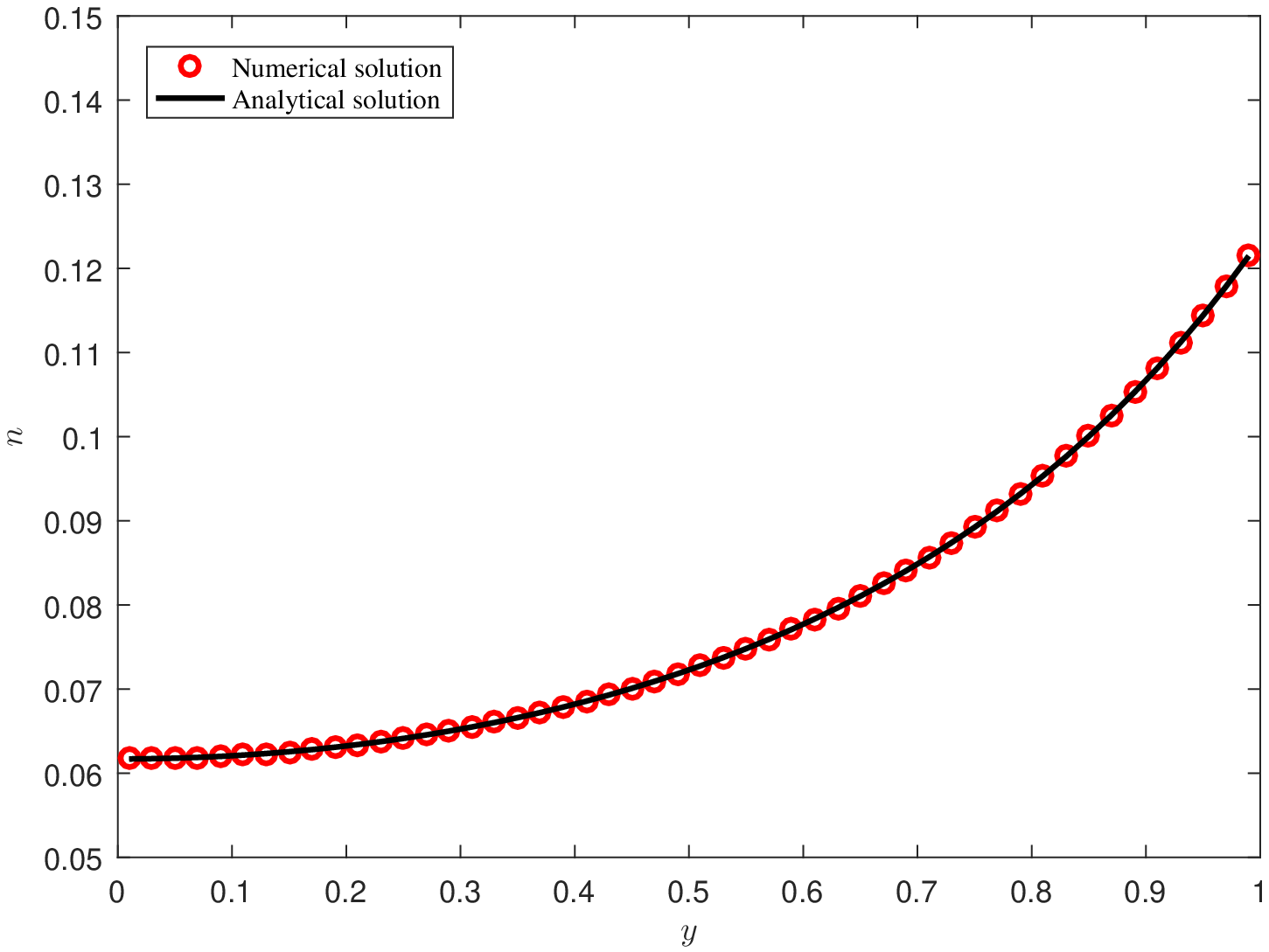}
			\label{fig-vn}
	\end{minipage}}
	\subfigure[]{
		\begin{minipage}{0.5\linewidth}
			\centering
			\includegraphics[width=2.5in]{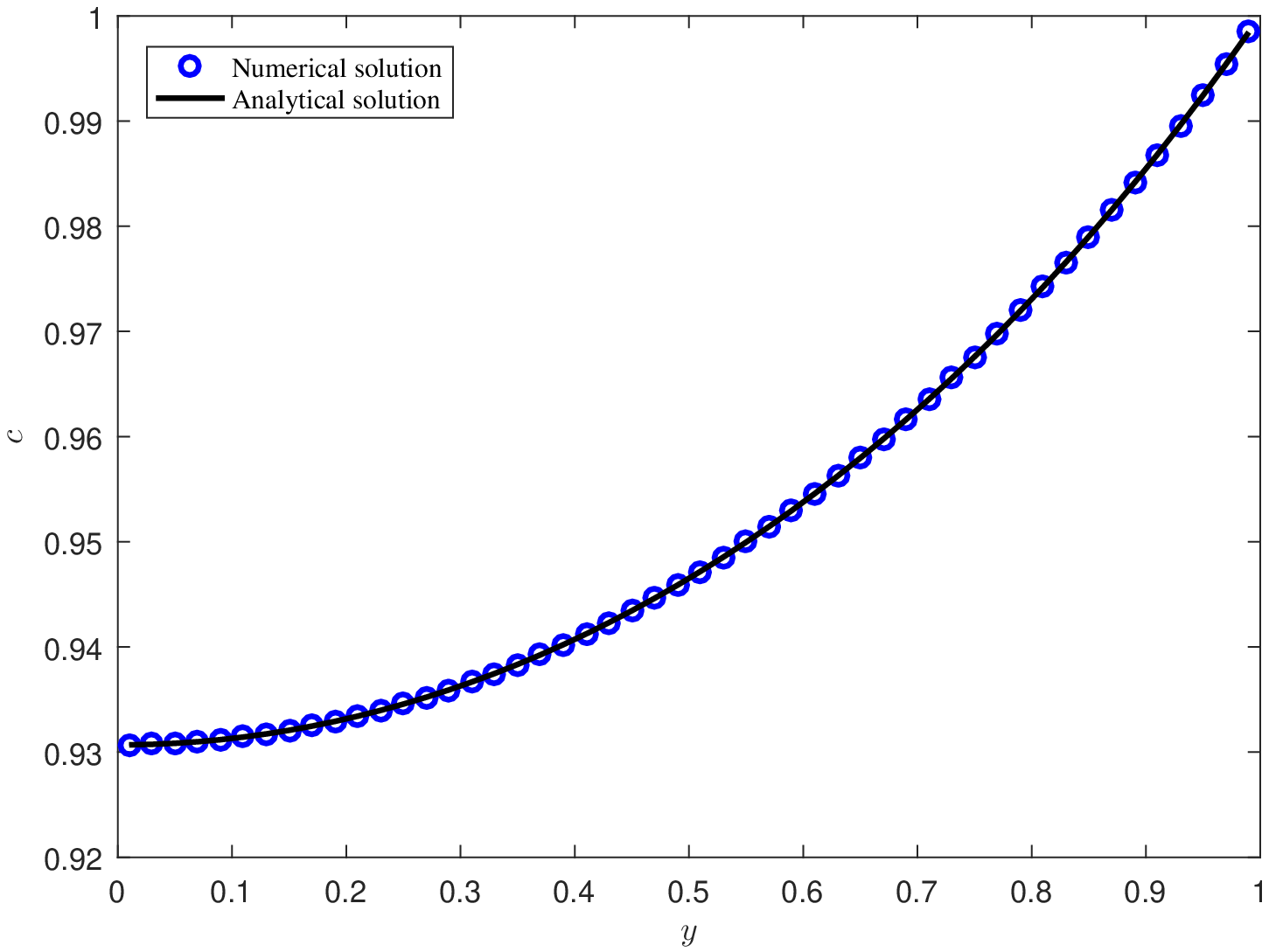}
			\label{fig-vc}
	\end{minipage}}
	\caption{A comparison of the numerical and analytical solutions [(a) $n$ and (b) $c$].}
	\label{fig-verify}
\end{figure}
Based on the results in Ref. \cite{Hillesdon1995BMB}, under some suitable parameters, the solutions of Eqs.\,(\ref{non-n})-(\ref{non-NS21}) converge to homogeneous-in-$x$ steady-state solutions of the following time-independent system:
\begin{equation}
\nabla\cdot[\nabla n-\alpha r(c)n\nabla c]=0,\quad \nabla\cdot\delta\nabla c-\beta r(c)n=0.
\end{equation} 
 The analytical solutions $n^{s}(y)$ and $c^{s}(y)$ can be explicitly derived if $c\geqslant c^{*}$ (i.e., $r(c)=1$ in the entire domain),
 \begin{equation}
 n^{s}(y)=\frac{\delta A^{2}}{\beta}\frac{\alpha}{2}\frac{1}{\cos^{2}(\frac{\alpha}{2}Ay)},\quad c^{s}(y)=1-\frac{2}{\alpha}\ln\left(\frac{\cos(\frac{\alpha}{2}Ay)}{\cos(\frac{\alpha}{2}A)}\right),
 \end{equation}
 where $A$ is a positive constant and satisfies
 \begin{equation}\label{A-relationship}
 \beta \int_{0}^{d} n^{s}(y)dy=\delta A\tan(\frac{\alpha}{2}A).
 \end{equation} 
 We now numerically study the system (\ref{non-KS-NS}) in the domain $\Omega=[-3, 3]\times [0, 1]$ with $\alpha=10$, $\beta=10$, $\delta=5$, $\gamma=1000$, $Sc=500$ and the following initial conditions,
 \begin{equation}
 n_{0}(x,y)=\frac{\pi}{40}, \quad c_{0}(x,y)=1, \quad \mathbf{u}_{0}(x,y)=\mathbf{0}. 
 \end{equation}
 According to Eq.\,(\ref{A-relationship}), one can determine $A=\pi/20$.
 In our simulations, we take $\Delta x=\Delta y=0.02$ and $\Delta t= 2.5\times 10^{-7}$ such that the relaxation factor is in a proper range. In order to determine whether the result reaches a stable state, the following criterion is adopted,
 \begin{equation}
 \frac{\sum_{i,j}\left(n^{N+1000}_{i,j}-n^{N}_{i,j}\right)}{\sum_{i,j}n^{N+1000}_{i,j}}<1\times 10^{-7},\quad  \frac{\sum_{i,j}\left(c^{N+1000}_{i,j}-c^{N}_{i,j}\right)}{\sum_{i,j}c^{N+1000}_{i,j}}<1\times 10^{-7},
 \end{equation}
 where $n^{N}_{i,j}$ and $c^{N}_{i,j}$ denote the bacteria and oxygen concentrations at position $(i\Delta x, j \Delta y)$ and time $N \Delta t$. 
 From Fig. \ref{fig-verify}, one can find the numerical results are in good agreement with the analytical solutions, and the relative errors of bacteria and oxygen concentrations are less than $1\times10^{-7}$.
 
 \subsubsection{The influences of some physical parameters}\label{parameter-influence}
 In this part, we consider an example with $\alpha =5$, $\beta = 5$, $\delta =0.25$, $\gamma =418$ and $Sc=7700$. The time step is adjusted to $\Delta t=1\times 10^{-8}$ due to the larger value of $Sc$. The cut-off value $c^{*}=0.3$, while the truncated function $r(c)$ is regularized as a continuously form,
 \begin{equation}
 r(c)=\frac{1}{2}\Big[1+\frac{c-c^{*}}{\sqrt{(c-c^{*})^{2}+\epsilon^{2}}}\Big],
 \end{equation}
 where $\epsilon$ is a positive constant close to zero, and is set as $\Delta x$ below.
    \begin{figure}	
 	\setlength{\abovecaptionskip}{-0.05cm} 
 	\setlength{\belowcaptionskip}{-0.3cm} 
 	\subfigure[]{
 		\begin{minipage}{0.5\linewidth}
 			\centering
 			\includegraphics[width=2.5in]{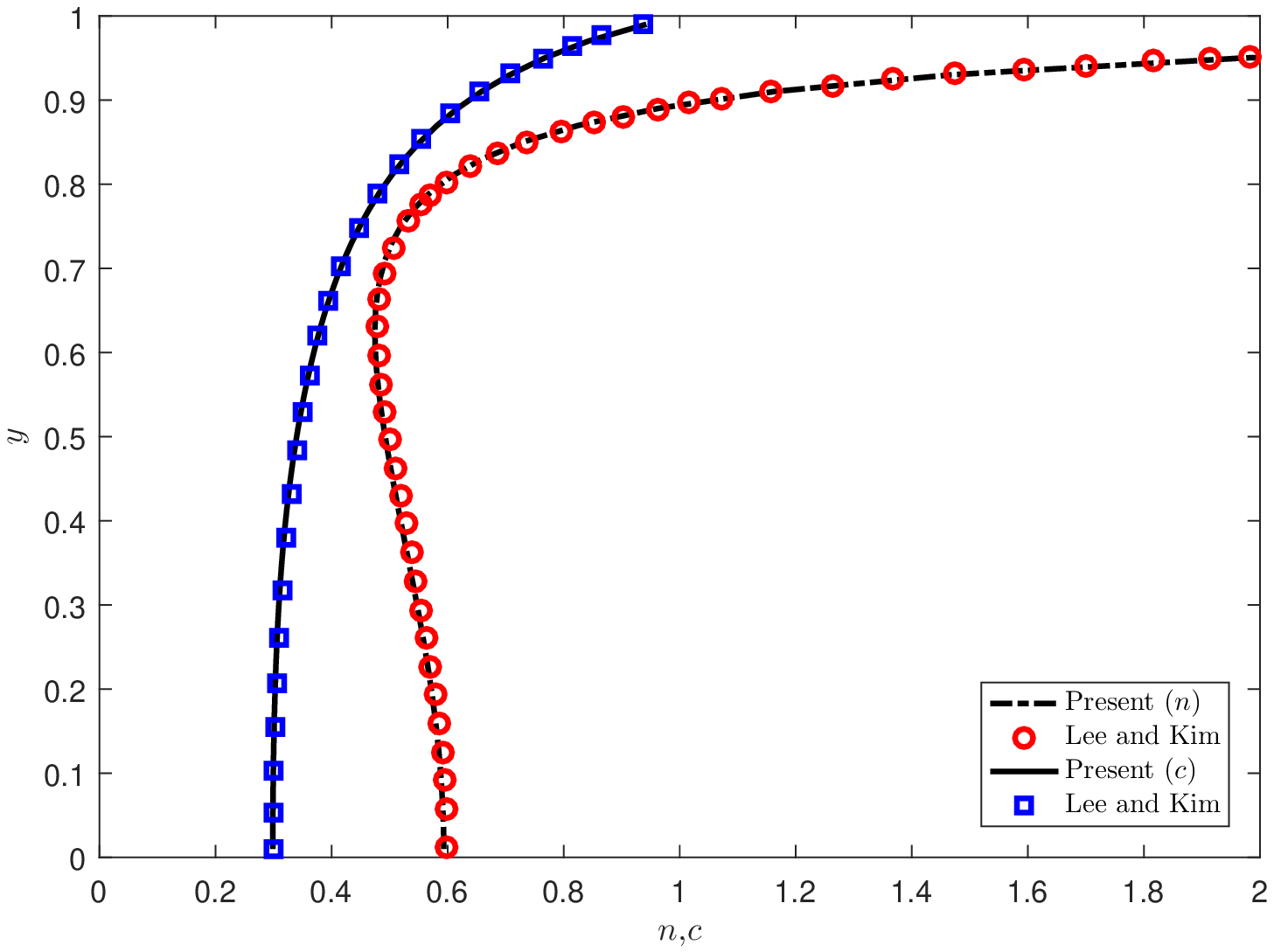}
 			\label{fig-vnc}
 	\end{minipage}}
 	\subfigure[]{
 		\begin{minipage}{0.5\linewidth}
 			\centering
 			\includegraphics[width=2.5in]{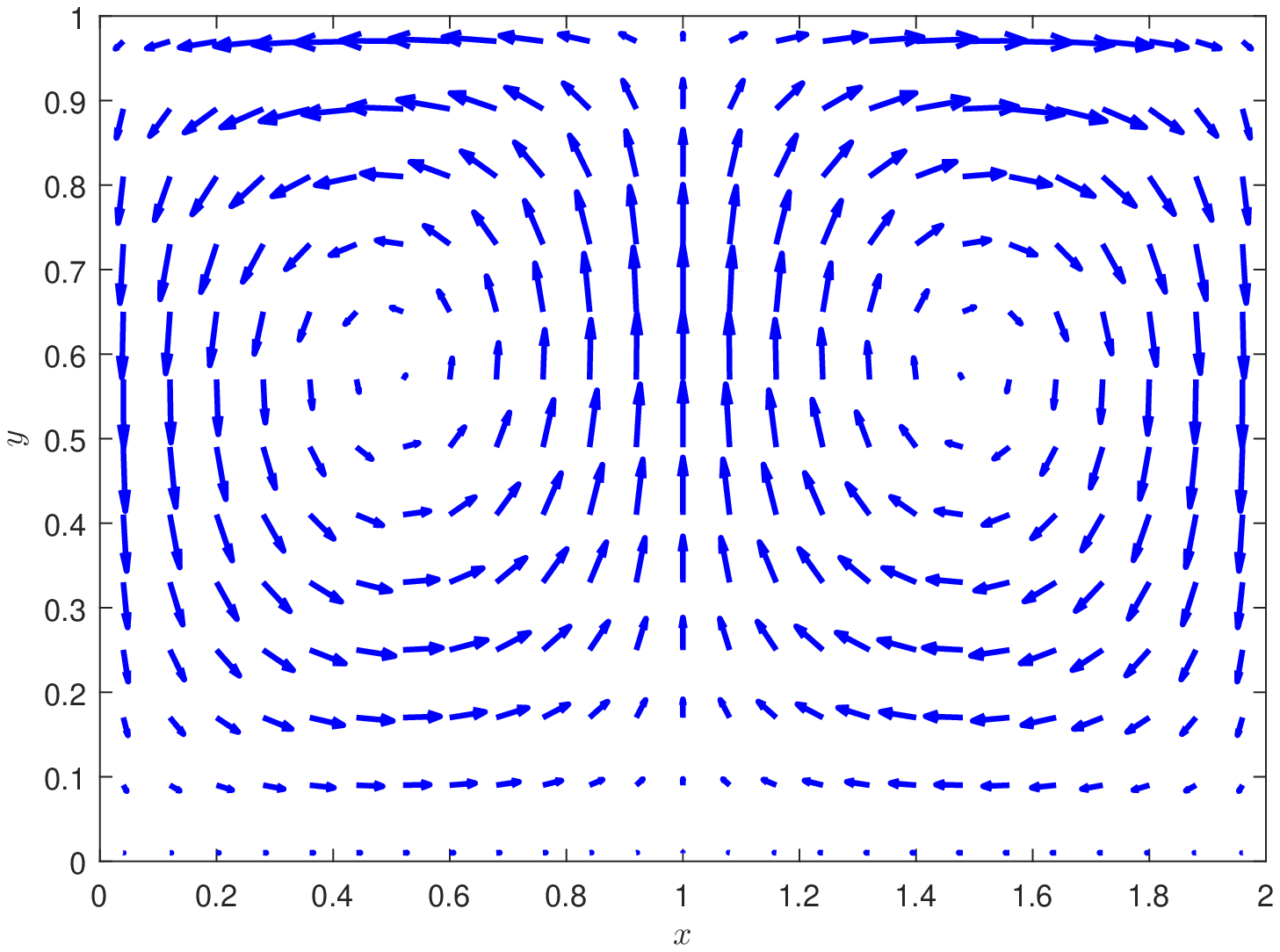}
 			\label{fig-quvier}
 	\end{minipage}}
 	\caption{(a) Vertical profiles of the bacteria and oxygen concentrations $n$ and $c$, (b) velocity vector.}
 	\label{fig-verify2}
 \end{figure}
 The initial conditions are given by
 \begin{equation}
 \begin{aligned}
 n_{0}(x,y)=\begin{cases}
 1& \text{if }y>0.501-0.01\sin((x-0.5)\pi),\\
 0.5& \text{otherwise,}
 \end{cases}\quad
 c_{0}(x,y)=1,\quad \mathbf{u}_{0}(x,y)=\mathbf{0}.
 \end{aligned}
 \end{equation}  
 This problem is used to validate the present LB model through a comparison with the results reported by Lee and Kim \cite{Lee2015EJMBF}. To this end, we present the quasi-homogeneous-in-$x$ vertical profiles of bacteria and oxygen concentrations at $t=0.22$ in Fig. \ref{fig-vnc}. The results in this figure shows that the bacteria increase towards the bottom of domain, this is because the chemotactic convection is cut-off for oxygen levels below $c^{*}$. In addition, the convection structure can be captured by the velocity field in Fig. \ref{fig-quvier}. 

\begin{figure}
 	\setlength{\abovecaptionskip}{-0.05cm} 
 	\setlength{\belowcaptionskip}{-0.3cm} 
 	\subfigure[]{
 		\begin{minipage}{0.5\linewidth}
 			\centering
 			\includegraphics[width=2.5in]{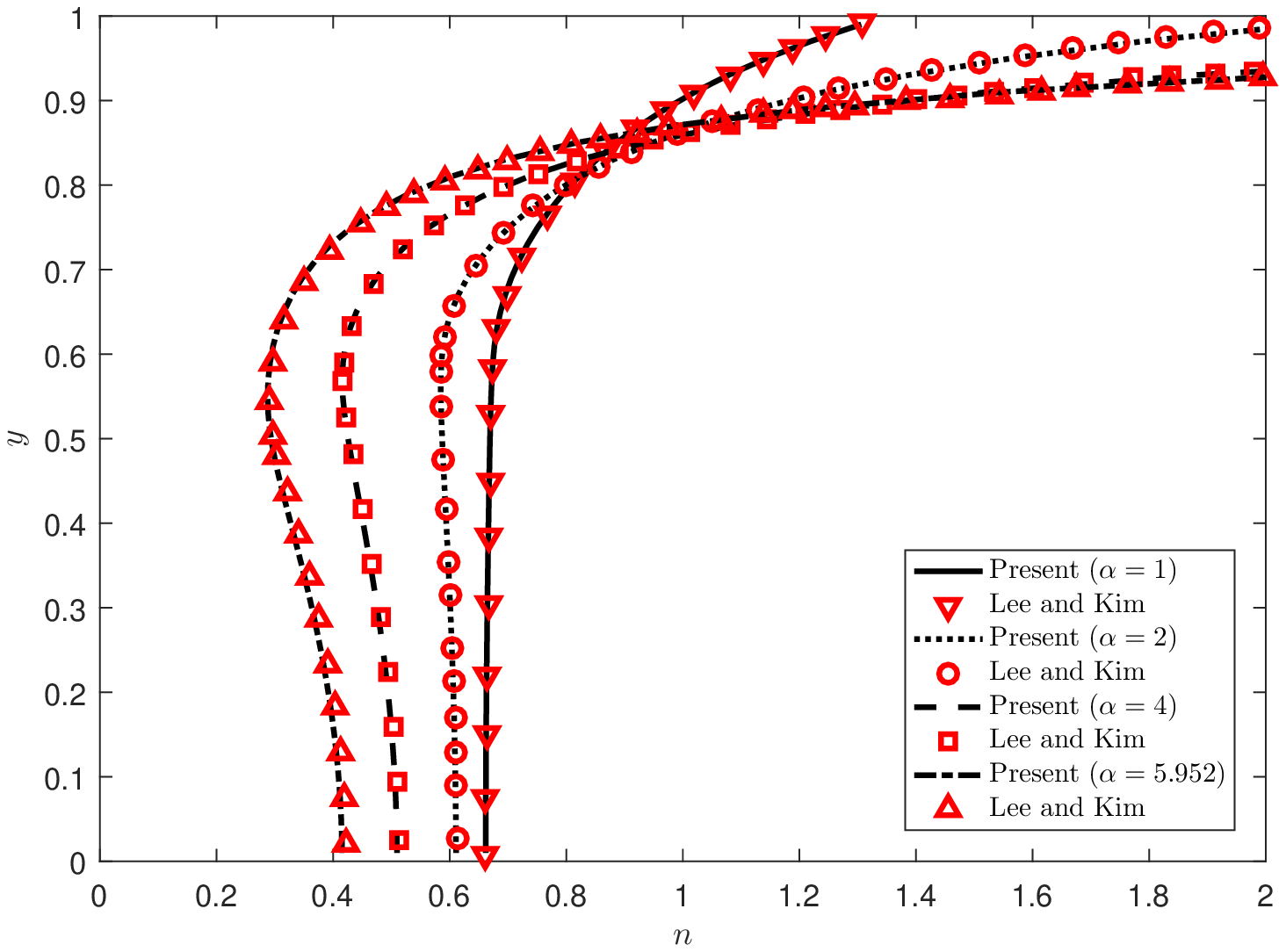}
 			\label{fig-alpha-n}
 	\end{minipage}}
 	\subfigure[]{
 		\begin{minipage}{0.5\linewidth}
 			\centering
 			\includegraphics[width=2.5in]{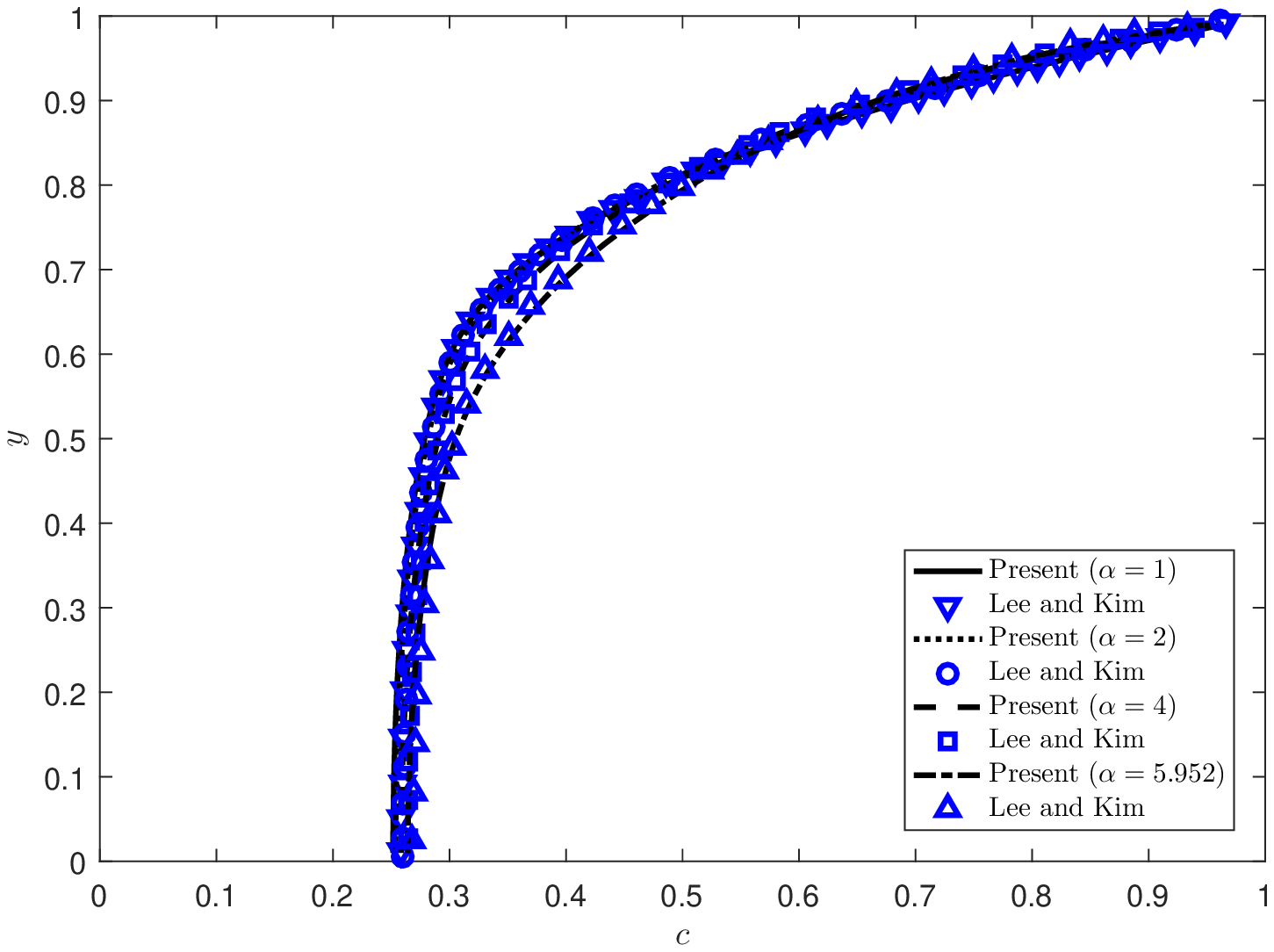}
 			\label{fig-alpha-c}
 	\end{minipage}}
 	\caption{Vertical profiles of the bacteria and oxygen concentrations [(a)\,$n$ and (b)\,$c$] at $t=0.22$ for $\alpha=1, 2, 4, 5.952$.}
 	\label{fig-alpha}
 \end{figure}

 Now we discuss the influence of parameter $\alpha$ to the system. Increasing $\alpha$ denotes the increase of bacteria directional swimming relative to their diffusion ($\beta$ and $\delta$ are fixed). For this purpose, we fixed the other parameters to be $\beta=10$ and $\delta=1$, and consider different values of  $\alpha$ ($\alpha=1, 2, 4, 5.952$). From Fig. \ref{fig-alpha} where the vertical profiles of bacteria, oxygen concentrations at $t=0.22$ are shown, one can find that with the increase of $\alpha$, the bacteria concentration increases near the surface of the domain, bacteria leave the lower part of the domain faster, and consume little oxygen. 
 
\begin{figure}
 	\setlength{\abovecaptionskip}{-0.05cm} 
 	\setlength{\belowcaptionskip}{-0.3cm} 
 	\subfigure[]{
 		\begin{minipage}{0.5\linewidth}
 			\centering
 			\includegraphics[width=2.5in]{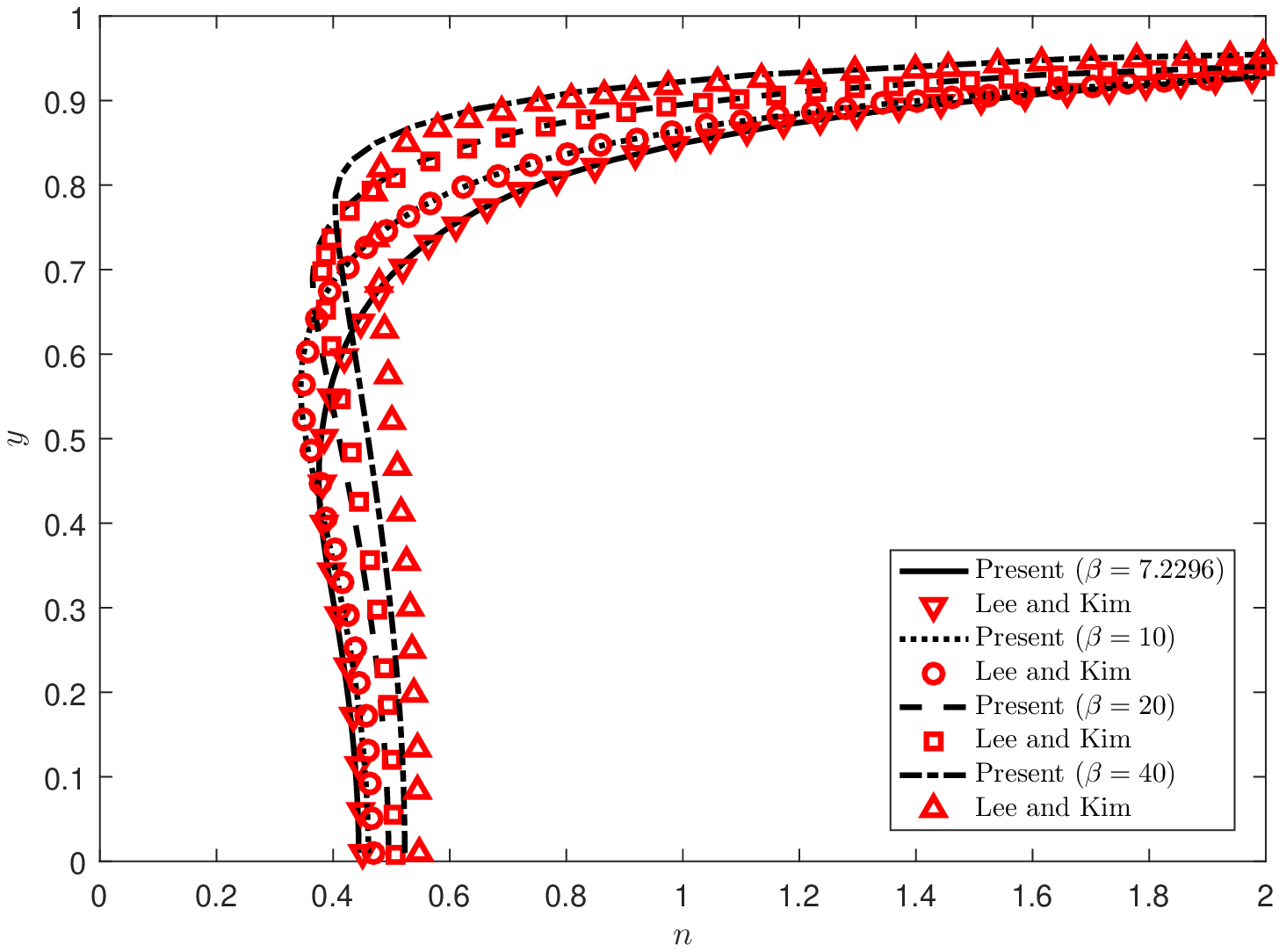}
 			\label{fig-beta-n}
 	\end{minipage}}
 	\subfigure[]{
 		\begin{minipage}{0.5\linewidth}
 			\centering
 			\includegraphics[width=2.5in]{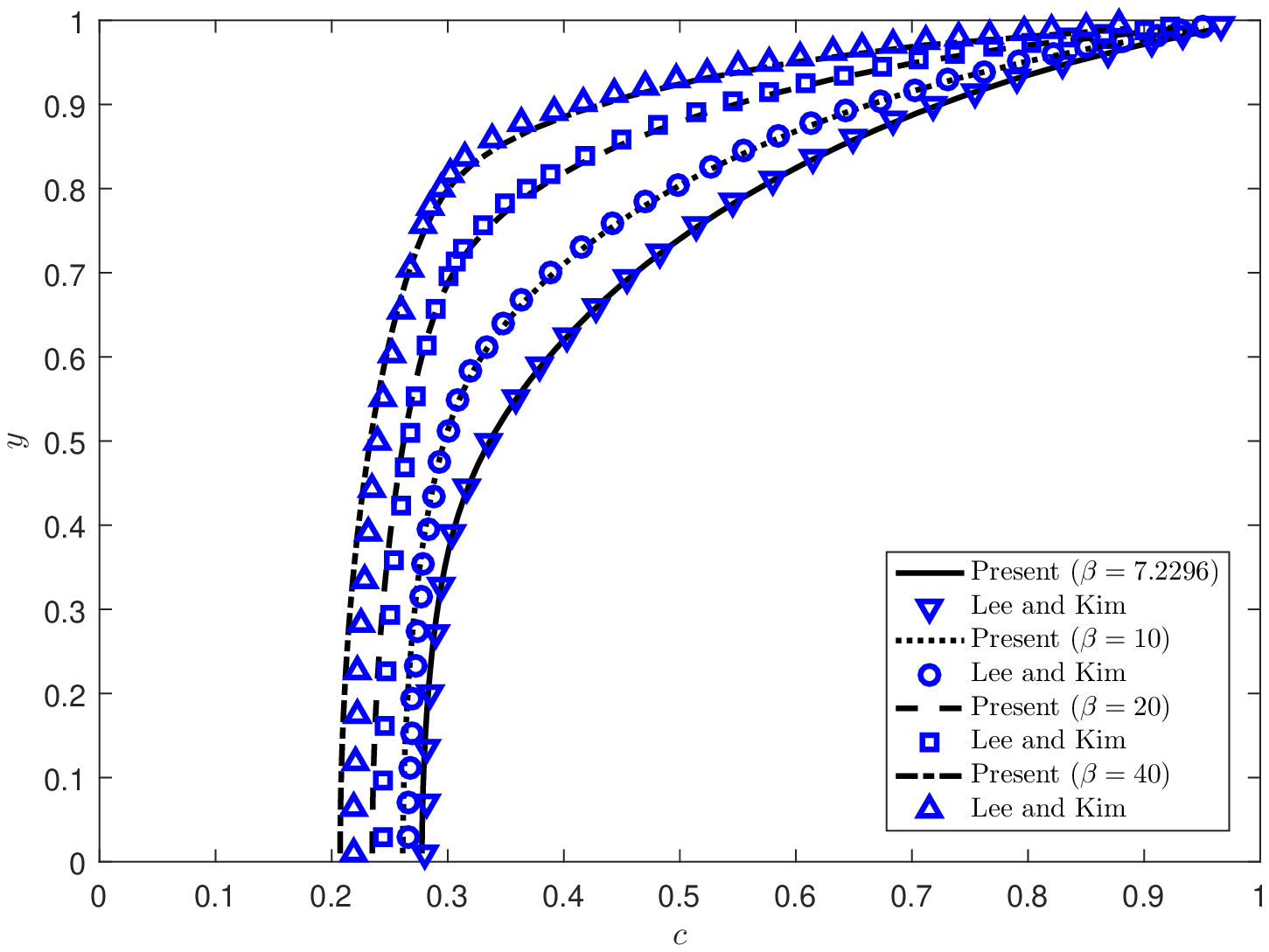}
 			\label{fig-beta-c}
 	\end{minipage}}
 	\caption{Vertical profiles of the bacteria and oxygen concentrations [(a)\,$n$ and (b)\,$c$] at $t=0.22$ for $\beta=7.2296, 10, 20, 40$.}
 	\label{fig-beta}
 \end{figure}

Then we consider the effect of parameter $\beta$. The increase of $\beta$ indicates the increase of oxygen consumption compared with oxygen diffusion ($\alpha$ and $\delta$ are fixed). As shown in Fig. \ref{fig-beta} where $\alpha=5$, $\delta=1$, $\beta = 7.2296, 10, 20$ and $40$ at $t=0.22$, with the increase of $\beta$, the oxygen concentration decreases at the same height, and the up-swimming bacteria increase. We note that the present results of different cases are in agreement with the previous work \cite{Lee2015EJMBF}.
 
\begin{figure}
 	\setlength{\abovecaptionskip}{-0.05cm} 
 	\setlength{\belowcaptionskip}{-0.3cm} 
 	\subfigure[]{
 		\begin{minipage}{0.5\linewidth}
 			\centering
 			\includegraphics[width=2.5in]{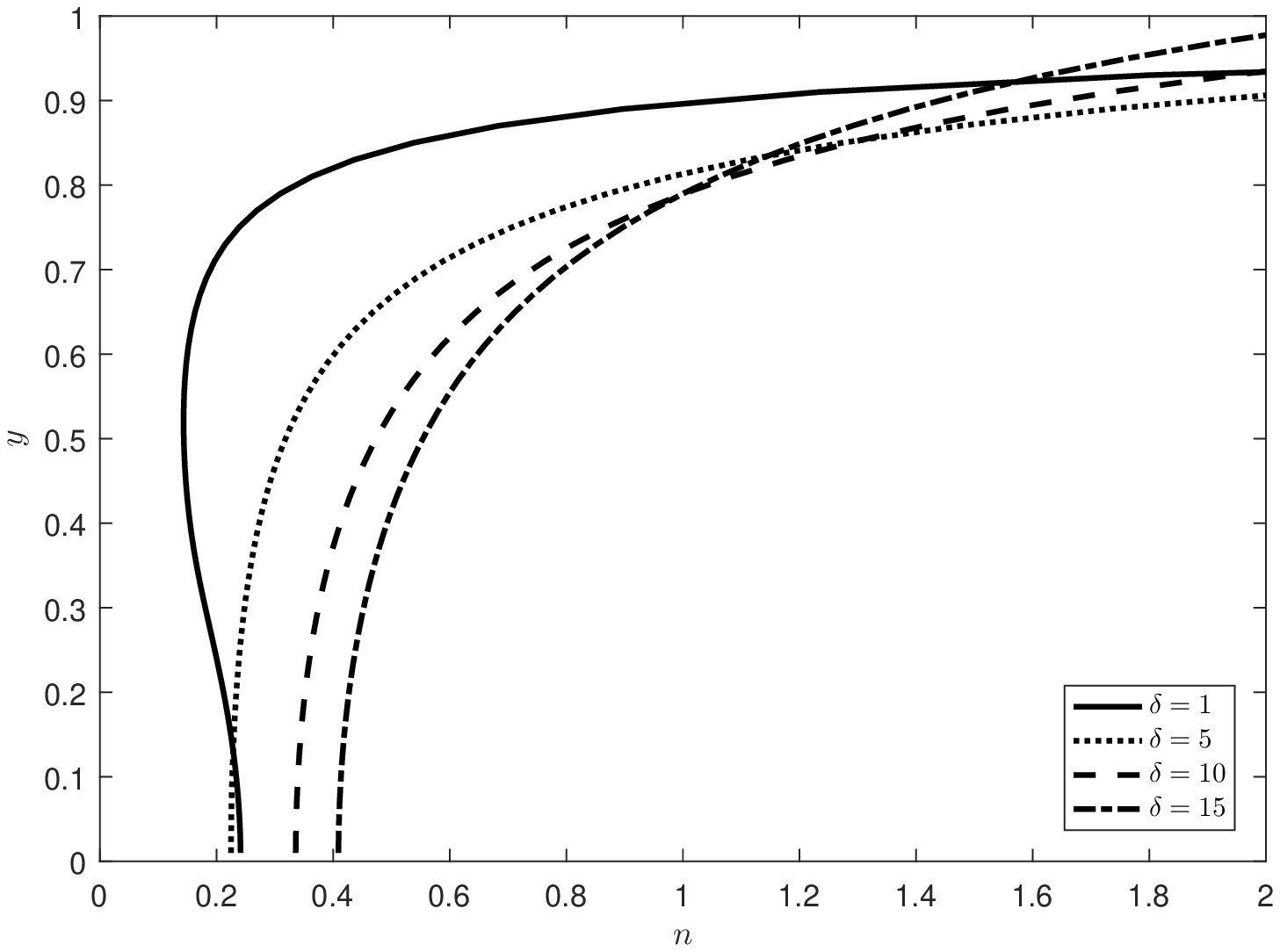}
 			\label{fig-delta-n}
 	\end{minipage}}
 	\subfigure[]{
 		\begin{minipage}{0.5\linewidth}
 			\centering
 			\includegraphics[width=2.5in]{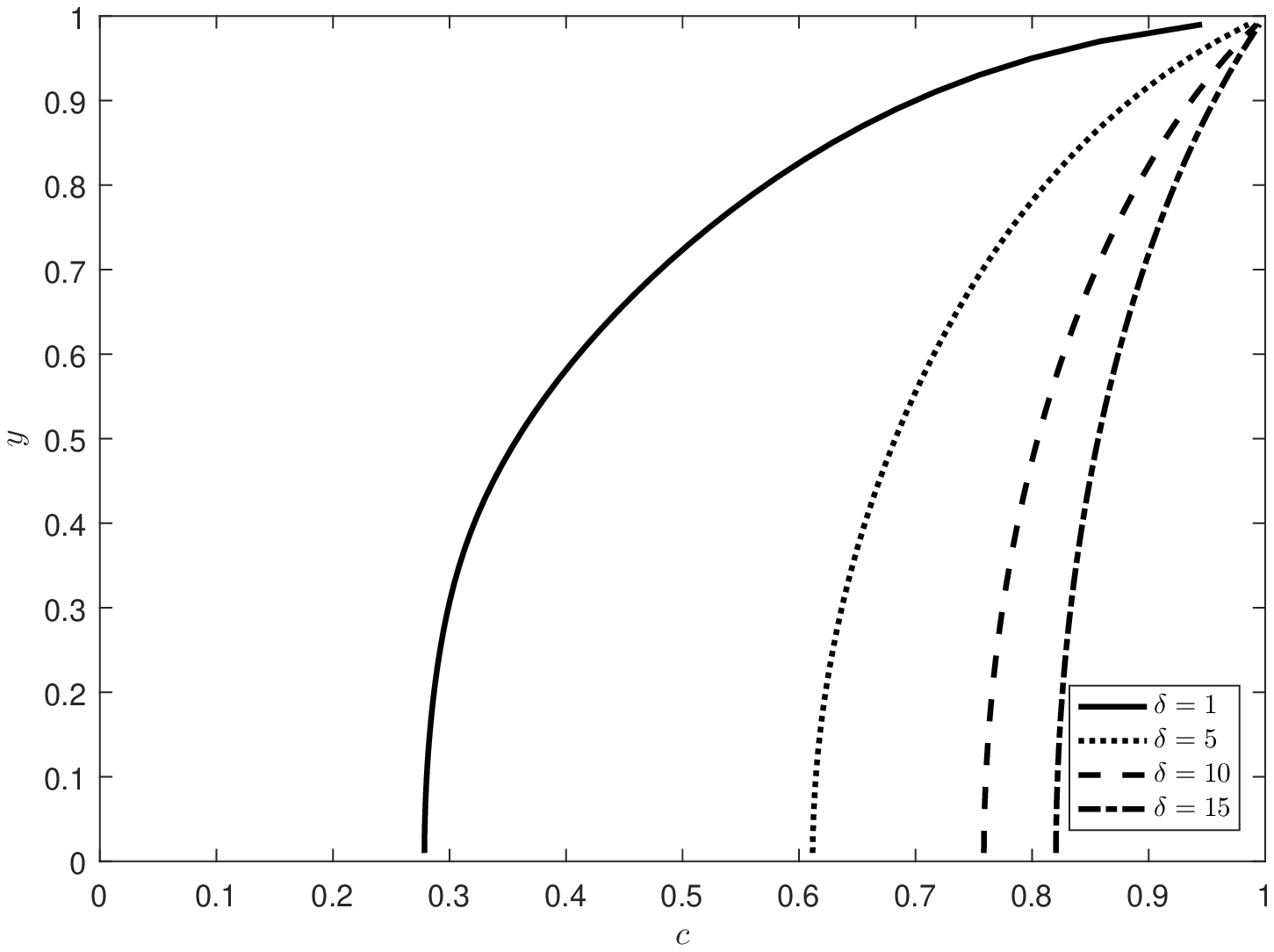}
 			\label{fig-delta-c}
 	\end{minipage}}
 	\caption{Vertical profiles of the bacteria and oxygen concentrations [(a)\,$n$ and (b)\,$c$] at $t=0.22$ for $\delta=1, 5, 10, 25$.}
 	\label{fig-delta}
 \end{figure}

Finally, we focus on the influence of parameter $\delta$. Actually, the increase of $\delta$ indicates that oxygen diffusion increases and oxygen adds to entire domain faster ($\alpha$ and $\beta$ are fixed). We take $\alpha=10$, $\beta=10$, $\delta=1, 5, 10, 15$, and present the vertical profiles of bacteria, oxygen concentrations at $t=0.22$ in Fig. \ref{fig-delta}, from which we can see that with the increase of $\delta$, oxygen diffuses into the entire domain faster, while most of bacteria stay at the lower part of the domain and the up-swimming bacteria decrease. 
 \subsubsection{The formation of plume structures} \label{plumes}
 In this part, we investigate the formation of plume structures caused by the high bacteria concentration near the surface under some certain parameters. In the following simulations, we take the bacteria-typical parameters as $\alpha =10$, $\delta=5$, $Sc=500$, and time step  $\Delta t=2.5\times 10^{-7}$. The truncated function is given by
 \begin{equation}
 r(c)=\begin{cases}
 1& \text{if }c\geqslant 0.3,\\
 0& \text{if }c<0.3.
 \end{cases}
 \end{equation}
  \begin{figure}
 	\vspace{-0.4cm} 
 	\setlength{\abovecaptionskip}{-0.05cm} 
 	\setlength{\belowcaptionskip}{-0.3cm} 
 	\centering
 	\includegraphics[width=5.5in]{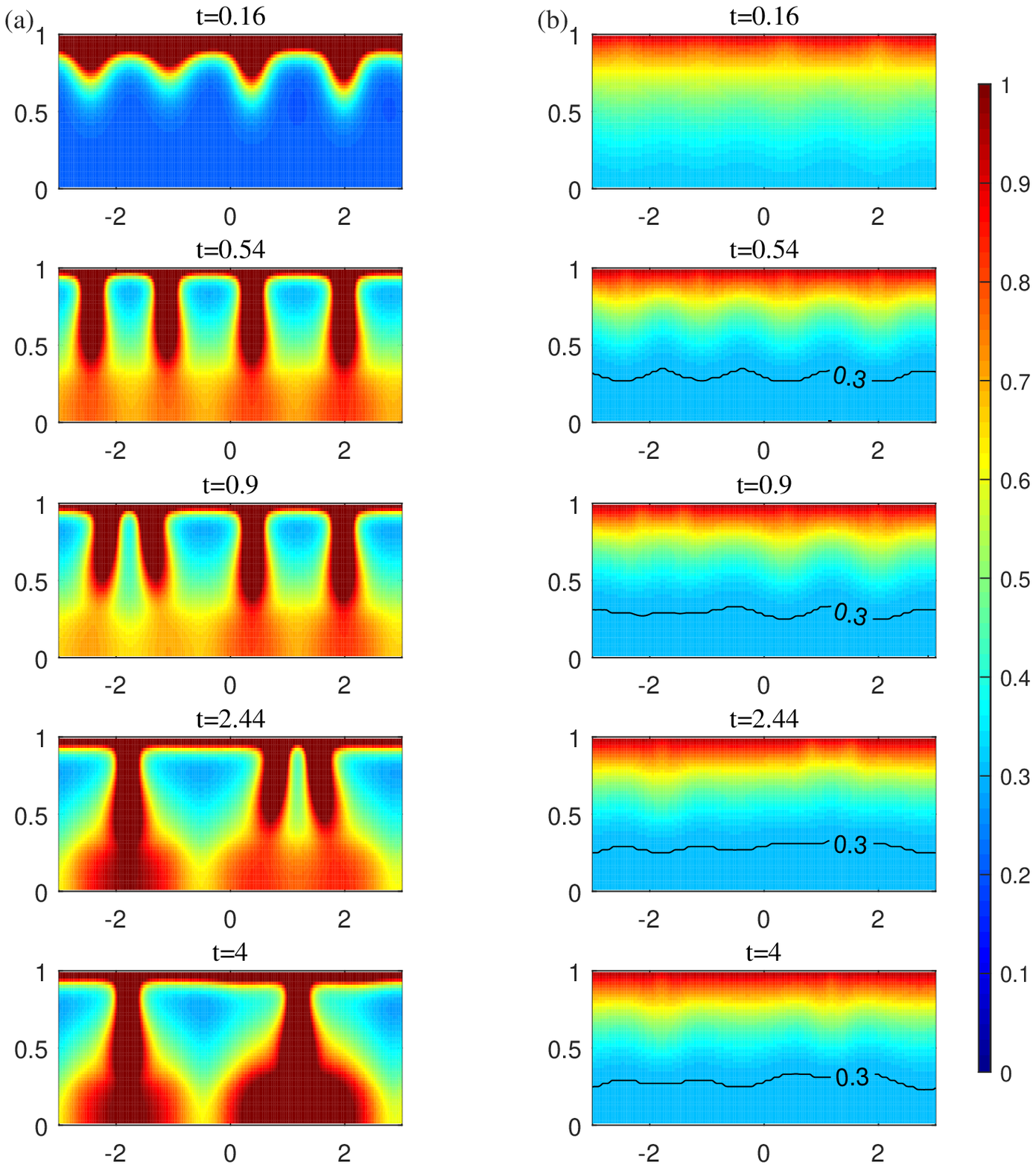}
 	\caption{Evolution of (a)\,bacteria concentration $n$ and (b)\,oxygen concentration $c$ in time.}
 	\label{fig-rand}
 \end{figure}
 \begin{figure}
 	\vspace{-0.4cm} 
 	\setlength{\abovecaptionskip}{-0.05cm} 
 	\setlength{\belowcaptionskip}{-0.5cm} 
 	\centering
 	\includegraphics[width=5.4in]{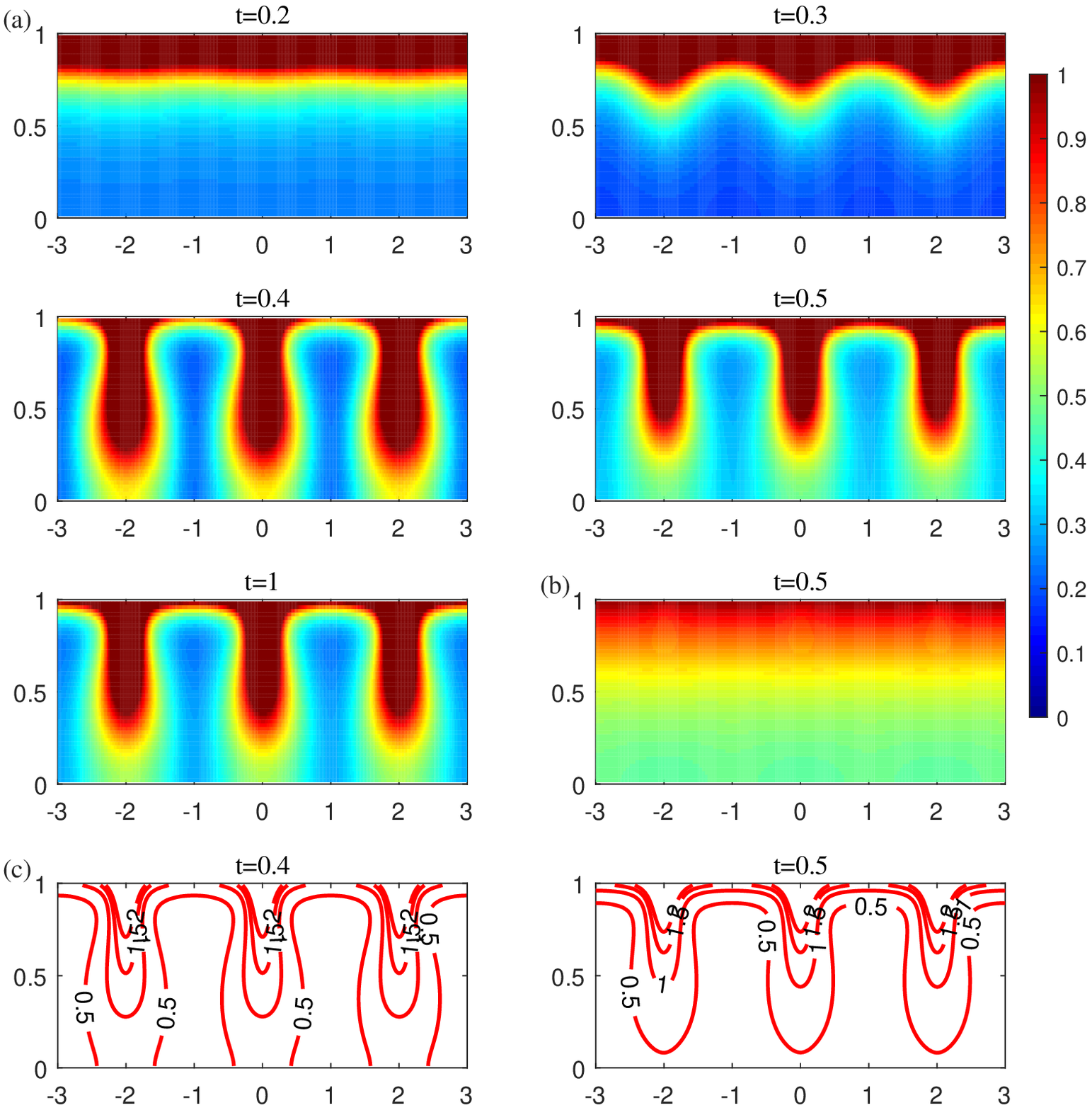}
 	\caption{(a)\,Evolution of bacteria concentration $n$ in time, (b)\,oxygen concentration $c$ at $t=0.5$, (c)\,level sets of bacteria concentration $n$ at $t=0.4, 0.5$.}
 	\label{fig-sin}
 \end{figure}
 
 We first consider the case with $\beta =20$ and $\gamma =2000$, and the following homogeneous initial data with a random perturbation in the bacteria concentration,
 \begin{equation}\label{rand-init}
 n_{0}(x,y)=0.8+0.2\xi,\quad c_{0}(x,y)=1,\quad \mathbf{u}_{0}(x,y)=\mathbf{0},
 \end{equation}
 where $\xi$ is a random number uniformly distributed in the interval $[0,1]$. Fig. \ref{fig-rand} shows the evolution of bacteria concentration $n$ and oxygen concentration $c$ in time. From this figure, we can observe the instability phenomena at $t=0.16$ due to the large amount of bacteria. Then the instability amplifies the random irregularity of initial data, and develops into four plumes at $t=0.54$. The oxygen concentration $c$ in most of the bottom half of the domain is less than the chemotaxis cut-off $c^{*}=0.3$ owing to the large amount of bacteria consumption, thus the bacteria become inactive and stop directed swimming in this region, just sink down into the bottom of the domain along with the fluid. Later at $t=0.9$, the left two plumes are close to each other, and merge into a large plume at $t=2.44$. Meanwhile, the two plumes at the right side also approach to each other, and combine together at $t=4$, this structure no longer changes in time.

 Then we cut characteristic bacteria density $n_{r}$ by half which means $\beta =10$ and $\gamma =1000$, and the deterministic initial data are given with small sinusoidal modulations,
 \begin{equation}\label{sin-data}
 \begin{aligned}
 n_{0}(x,y)=\begin{cases}
 1& \text{if }y>0.499-0.01\sin((x-1.5)\pi),\\
 0.5& \text{otherwise,}
 \end{cases}\quad c_{0}(x,y)=1,\quad \mathbf{u}_{0}(x,y)=\mathbf{0}.
 \end{aligned}
 \end{equation}
 We carry out some simulations, and present the results at different times in Fig. \ref{fig-sin}. From Fig. \ref{fig-sin}(a), one can observe that the instability occurs at the lower edge of the high concentration layer at $t=0.2$. When the time is increased to $t=0.3$, three plumes are formed due to the modulations in initial conditions (\ref{sin-data}). It is interesting that those plumes bounce upwards slightly at $t=0.5$ after arriving the bottom of the domain at $t=0.4$, which can be seen from Fig. \ref{fig-sin}(c). The plumes have some minor changes, and continue dropping to the bottom of the domain slightly from $t=0.5$ to $t=1.0$. In addition, we also plot the oxygen concentration at $t=0.5$ in Fig. \ref{fig-sin}(b), which confirms that the oxygen level remains above the cut-off value $c^{*}=0.3$ in the whole domain. We note that these results agree with those in Ref. \cite{Chertock2012JFM}.
 
\subsection{Numerical results and discussion on the double-diffusive convection system}
Similar to above discussion, we can derive the DDC system with Soret and Dufour effects from CDF system by setting  $\mathbf{\Phi}=(T,C)^{T}$, $\mathbf{D}=\begin{bmatrix}
\alpha & k_{CT}\\
k_{TC} & D	
\end{bmatrix}$, $\mathbf{S}=\mathbf{0}$, $\mathbf{F}=\mathbf{g}[1-\beta_{T}(T-T_{0})-\beta_{C}(C-C_{0})]$ and $p=p/\rho_{0}$,
\begin{subequations}\label{DS-NS}
	\begin{equation}\label{DS-T}
	T_{t}+\nabla\cdot T\mathbf{u}=\nabla\cdot[\alpha\nabla T+k_{CT}\nabla C],
	\end{equation}
	\begin{equation}\label{DS-C}
	C_{t}+\nabla\cdot C\mathbf{u}=\nabla\cdot[k_{TC}\nabla T+D\nabla C],
	\end{equation}
	\begin{equation}\label{DS-NS1}
	\nabla\cdot\mathbf{u}=0,
	\end{equation}
	\begin{equation}\label{DS-NS2}
	\mathbf{u}_{t}+\mathbf{u}\cdot\nabla\mathbf{u}=-\nabla p/\rho_{0}+\nabla\cdot\nu\nabla\mathbf{u}+\mathbf{g}[1-\beta_{T}(T-T_{0})-\beta_{C}(C-C_{0})],
	\end{equation}	
\end{subequations}
 where $T$ is the temperature and $C$ is the concentration. $\alpha$ and $D$ are the thermal diffusivity and mass diffusivity. $k_{CT}$ and $k_{TC}$ are the Dufour and Soret coefficients, respectively. In Eq.\,(\ref{DS-NS2}), the Boussinesq approximation is adopted here to consider the density in the buoyancy term with $T_{0}$ and $C_{0}$ being the reference temperature and concentration, $\beta_{T}$ and $\beta_{C}$ being the coefficients of thermal and solute expansion.

The boundary conditions of this system in 2D cavity $[0,L]\times[0,H]$ are given by
\begin{subequations}
	\begin{equation}
	\nabla T\cdot\hat{\mathbf{n}}=0,\quad \nabla C\cdot\hat{\mathbf{n}}=0,\quad \mathbf{u}=\mathbf{0},\quad  \text{at } y=0\text{ or }y=H,
	\end{equation}
	\begin{equation}
	T=T_{h},\quad C=C_{h},\quad \mathbf{u}=\mathbf{0},\quad \text{at } x=0,
	\end{equation}
	\begin{equation}
	T=T_{l},\quad C=C_{l},\quad \mathbf{u}=\mathbf{0},\quad \text{at } x=L,
	\end{equation}
\end{subequations}
where $T_{h}$ and $C_{h}$ are the higher temperature and concentration, while $T_{l}$ and $C_{l}$ are the lower ones. Additionally, the initial conditions are
\begin{equation}
T=T_{l},\quad C=C_{l},\quad \mathbf{u}=\mathbf{0},\quad \text{at } t=0.
\end{equation}

The governing equations (\ref{DS-T})-(\ref{DS-NS2}) can be expressed as a dimensionless form through introducing the following variables \cite{Ren2016IJHMT,Xu2019IJTS}:
\begin{equation}
\mathbf{x}'=\frac{\mathbf{x}}{L},\quad  t'=\frac{\alpha}{L^{2}}t,\quad  \mathbf{u}'=\frac{L\mathbf{u}}{\alpha},\quad  p'=\frac{L^{2}}{\alpha^{2}\rho_{0}}p,\quad 
T'=\frac{T-T_{0}}{T_{h}-T_{l}},\quad 
C'=\frac{C-C_{0}}{C_{h}-C_{l}}.                
\end{equation}
After dropping the prime notation in the rescaled variables, one can obtain the following dimensionless system:
\begin{subequations}\label{non-DS-NS}
	\begin{equation}
	T_{t}+\nabla\cdot T\mathbf{u}=\nabla\cdot[\nabla T+D_{CT}\nabla C],
	\end{equation}
	\begin{equation}
	C_{t}+\nabla\cdot C\mathbf{u}=\nabla\cdot[\frac{S_{TC}}{Le}\nabla T+\frac{1}{Le}\nabla C],
	\end{equation}
	\begin{equation}
	\nabla\cdot\mathbf{u}=0,
	\end{equation}
	\begin{equation}
	\mathbf{u}_{t}+\mathbf{u}\cdot\nabla\mathbf{u}=-\nabla p+\nabla\cdot Pr\nabla\mathbf{u}+Ra Pr(T+NcC)\hat{\mathbf{z}},
	\end{equation}
\end{subequations}
where Prandtl number $Pr$, Rayleigh number $Ra$, buoyancy ratio $Nc$, Lewis number $Le$, Dufour factor $D_{CT}$, Soret factor $S_{TC}$, and aspect ratio $A$  are defined as
\begin{equation}
\begin{aligned}
Pr=\frac{\nu}{\alpha},\quad &Ra=\frac{g\beta_{T}(T_{h}-T_{l})L^{3}}{\nu\alpha},\quad Nc=\frac{\beta_{C}(C_{h}-C_{l})}{\beta_{T}(T_{h}-T_{l})},\quad Le=\frac{\alpha}{D},\\
D_{CT}&=\frac{k_{CT}(C_{h}-C_{l})}{\alpha(T_{h}-T_{l})},\quad S_{TC}=\frac{k_{TC}(T_{h}-T_{l})}{D(C_{h}-C_{l})},\quad A=\frac{H}{L}.
\end{aligned}
\end{equation}

Similarly, we can also derive the dimensionless boundary and initial conditions,
\begin{subequations}\label{natural}
	\begin{equation}
	\nabla T\cdot\hat{\mathbf{n}},\quad \nabla C\cdot\hat{\mathbf{n}}=0,\quad \mathbf{u}=\mathbf{0},\quad \text{at } y=0\text{ or }y=A,
	\end{equation}
	\begin{equation}
	T=1,\quad C=1,\quad \mathbf{u}=\mathbf{0},\quad \text{at } x=0,
	\end{equation}
	\begin{equation}
	T=0,\quad C=0,\quad \mathbf{u}=\mathbf{0},\quad \text{at } x=1.
	\end{equation}
\end{subequations}
\begin{equation}
T=0,\quad C=0,\quad \mathbf{u}=\mathbf{0},\quad \text{at } t=0.
\end{equation}

The present LB model can also simulate above dimensionless system with $\mathbf{\Phi}=(T,C)^{T}$, $\mathbf{D}=\begin{bmatrix}
1 & D_{CT}\\
S_{TC}/Le & 1/Le	
\end{bmatrix}$, $\mathbf{S}=\mathbf{0}$, $\nu=Pr$ and  $\mathbf{F}=RaPr(T+NcC)\hat{\mathbf{z}}$.

To characterize the heat and mass transfer in DDC system, the local Nusselt number $Nu(y)$ and Sherwood number $Sh(y)$, and the average Nusselt number $\overline{Nu}$ and Sherwood number $\overline{Sh}$ on high temperature and concentration wall are defined as
\begin{subequations}
	\begin{equation}
	Nu(y)=\frac{\partial T(0,y)}{\partial x}+D_{CT}\frac{\partial C(0,y)}{\partial x},\quad \overline{Nu}=\frac{1}{A}\int_{0}^{A}Nu(y)dy,
	\end{equation}
	\begin{equation}
	Sh(y)= S_{TC}\frac{\partial T(0,y)}{\partial x}+\frac{\partial C(0,y)}{\partial x},\quad \overline{Sh}=\frac{1}{A}\int_{0}^{A}Sh(y)dy,
	\end{equation}
\end{subequations}
where the space derivative terms in $Nu(y)$ and $Sh(y)$ are calculated with the second-order upwind difference schemes. 

\begin{table}
	\centering
	\caption{A grid-independence study at $Ra=10^{5}$, $Pr=1.0$, $Le=2.0$, $A=2.0$, $Nc=-2.0$, $S_{TC}=0.1$ and $D_{CT}=0.1$.}
	\begin{tabular}{ccccc}
		\toprule
		Grid number & $\overline{Nu}$ & Deviation (\%) & $\overline{Sh}$ & Deviation (\%) \\
		\midrule
		$250\times250A$ & 2.9134 & - & 4.6891 & - \\
		$300\times300A$ & 2.9151 & 0.0583 & 4.6914 & 0.0490 \\
		$350\times350A$ & 2.9172 & 0.0720 & 4.6942 & 0.0596 \\
		\bottomrule
	\end{tabular}
	\label{table-grid}
\end{table}
\begin{table}
	\centering
	\caption{A comparison of $\overline{Nu}$ and $\overline{Sh}$ between the present work and Refs. \cite{Ren2016IJHMT,Xu2019IJTS} at $Pr=1.0$, $Le=2.0$, $Nc=-2.0$, $S_{TC}=0.1$ and $D_{CT}=0.1$.}
	\begin{tabular}{ccccccc}
		\toprule
		& \multicolumn{2}{c}{Present} & \multicolumn{2}{c}{Ref. \cite{Ren2016IJHMT}} & \multicolumn{2}{c}{Ref. \cite{Xu2019IJTS}} \\
		\cline{2-7}
		Rayleigh number $Ra$ & $\overline{Nu}$ & $\overline{Sh}$ & $\overline{Nu}$ & $\overline{Sh}$ & $\overline{Nu}$ & $\overline{Sh}$ \\
		\midrule
		$10^{3}$ & 1.2015 & 1.4694 & 1.2020 & 1.4683 & 1.2027 & 1.4654 \\
		$10^{4}$ & 1.7389 & 2.5923 & 1.7407 & 2.5943 & 1.7443 & 2.5766 \\
		$10^{5}$ & 2.9151 & 4.6914 & 2.9153 & 4.6993 & 2.8976 & 4.6108 \\
		$10^{6}$ & 5.2338 & 8.5781 & 5.1808 & 8.5134 & 5.0852 & 8.3294 \\
		$10^{7}$ & 9.4138 & 15.496 & 9.2426 & 15.210 &   -    &   -    \\
		\bottomrule
	\end{tabular}
	\label{table-dRa}
\end{table}
\subsubsection{A grid-independence study}
We first conduct a grid-independence study on DDC system in the cavity $[0,1]\times[0,A]$ at $Ra=10^{5}$, $Pr=1.0$, $Le=2.0$, $A=2.0$, $Nc=-2.0$, $S_{TC}=0.1$ and $D_{CT}=0.1$. The results of the average Nusselt number $\overline{Nu}$ and Sherwood number $\overline{Sh}$ at different grid numbers of $250\times250A$, $300\times300A$, and $350\times350A$ are presented in Table \ref{table-grid}. From this table, one can find that the relative deviations of $\overline{Nu}$ and $\overline{Sh}$ are less than 0.08\%. Based on these results, one can find that the grid size $300\times300A$ is fine enough, and is also used in the following simulations.

\begin{figure}
	\setlength{\abovecaptionskip}{-0.05cm} 
	\setlength{\belowcaptionskip}{-0.3cm} 
	\centering
	\subfigure[]{ 
		\begin{minipage}[]{1\linewidth} 
			\includegraphics[width=5.5in]{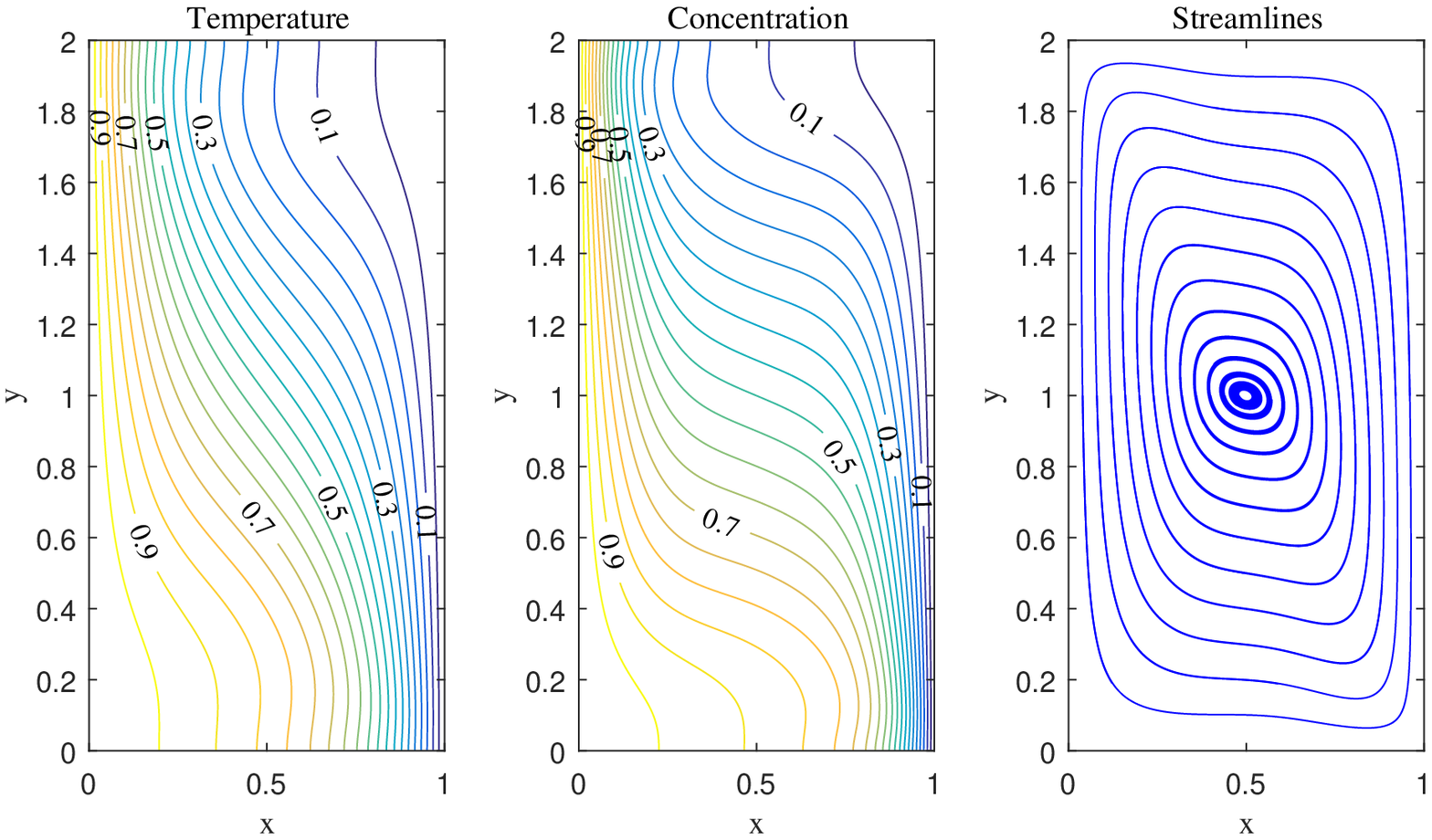}
			\centering
	\end{minipage}}
	\subfigure[]{ 
		\begin{minipage}[]{1\linewidth} 
			\includegraphics[width=5.5in]{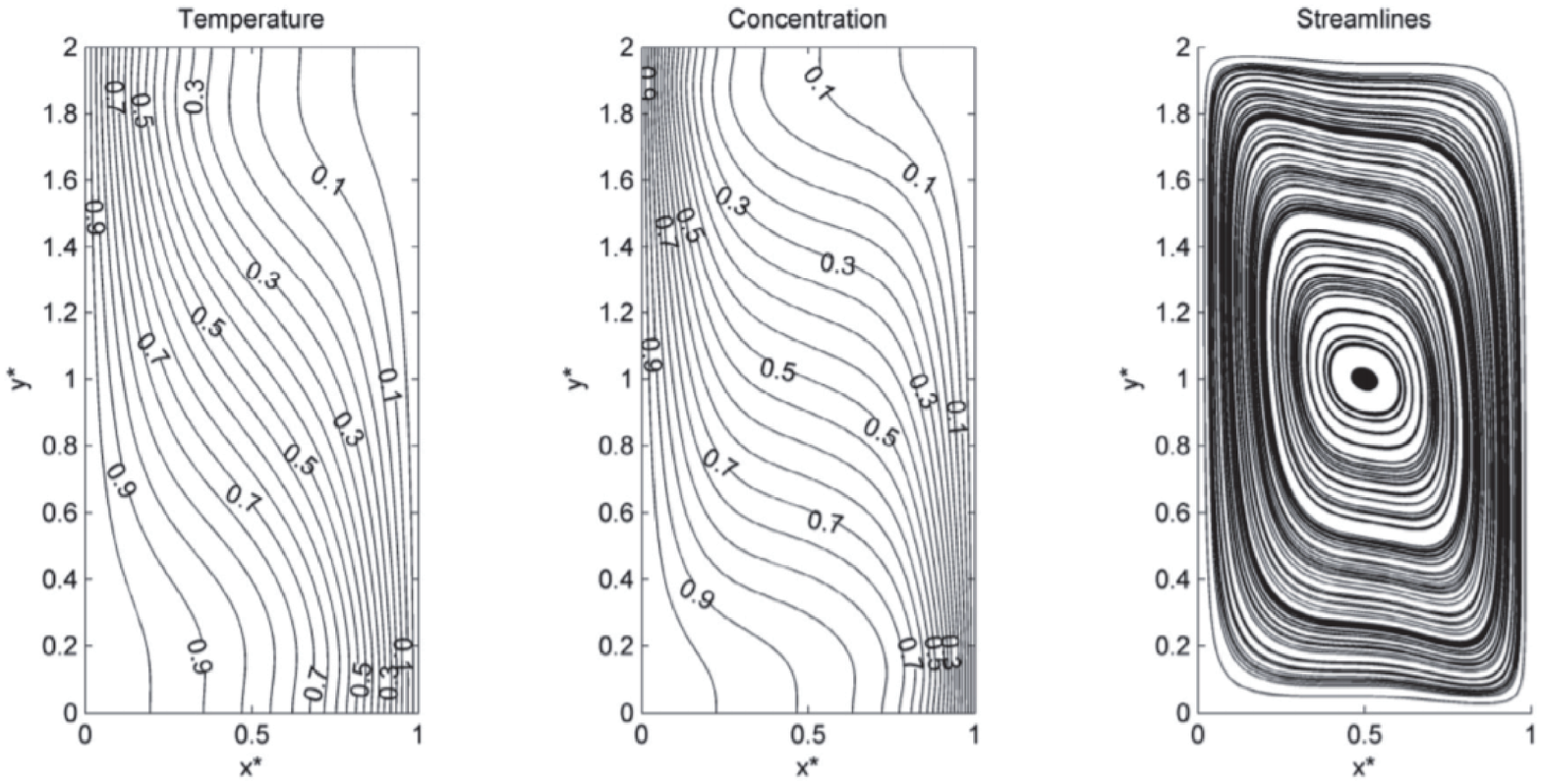}
			\centering
	\end{minipage}}
	\caption{Temperature, concentration and streamlines at $Ra=10^4$, $Pr=1.0$, $Le=2.0$, $Nc=-2.0$, $S_{TC}=0.1$ and $D_{CT}=0.1$. [(a) present results, (b) results in Ref. \cite{Ren2016IJHMT}].}
	\label{fig-Ra10^4}
\end{figure}

\subsubsection{The effects of some physical parameters}
We first consider the effect of Rayleigh number $Ra$. To this end, some other parameters are fixed as $Pr=1.0$, $Le=2.0$, $Nc=-2.0$, $A=2.0$, $S_{TC}=0.1$ and $D_{CT}=0.1$. As we can see from Table \ref{table-dRa}, the average Nusselt and Sherwood numbers increase with the increase of $Ra$. At $Le=2.0$, the concentration diffusion is weaker than the thermal diffusion, which gives a larger concentration gradient at boundary layer with a high concentration, thus the average Sherwood number is larger than the average Nusselt number. We note that present results are in good agreement with those in the previous works \cite{Ren2016IJHMT,Xu2019IJTS}, as shown in Table \ref{table-dRa} and Fig. \ref{fig-Ra10^4}, where the relative errors are no more than 2\%.

Then the effect of aspect ratio $A$ is also investigated. We perform some simulations at $Ra=10^{5}$, $Pr=1.0$, $Le=2.0$, $Nc=-2.0$, $S_{TC}=0.1$ and $D_{CT}=0.1$, and show the results in Table \ref{table-A} where the aspect ratio $A$ is changed from $1/6$ to $6$. From this table, one can see that the average Nusselt number increases when $A$ is increased up to $1.5$, and decreases when the aspect ratio is further increased, this means that there is a critical aspect ratio $A=1.5$. 
\begin{table}
	\centering
	\caption{The average Nusselt and Sherwood numbers of the DDC system at $Ra=10^{5}$, $Pr=1.0$, $Le=2.0$, $Nc=-2.0$, $S_{TC}=0.1$ and $D_{CT}=0.1$.}
	\begin{tabular}{ccccccc}
		\toprule
		& \multicolumn{2}{c}{Present} & \multicolumn{2}{c}{Ref. \cite{Ren2016IJHMT}} & \multicolumn{2}{c}{Deviation (\%)} \\
		\cline{2-7}
		Aspect ratio $A$ & $\overline{Nu}$ & $\overline{Sh}$ & $\overline{Nu}$ & $\overline{Sh}$ & $\overline{Nu}$ & $\overline{Sh}$ \\
		\midrule
		1/6 & 1.1108 & 1.1528 &   -    &    -   &    -    &    -    \\
		1/4 & 1.2582 & 1.7685 &   -    &    -   &    -    &    -    \\
		1/2 & 2.2893 & 4.3113 &   -    &    -   &    -    &    -    \\
		1   & 2.8925 & 4.9596 &   -    &    -   &    -    &    -    \\
		1.5 & 2.9590 & 4.8620 &   -    &    -   &    -    &    -    \\
		2   & 2.9151 & 4.6914 & 2.9153 & 4.6993 & 0.0069 & 0.1681 \\
		4   & 2.6508 & 4.1383 & 2.6505 & 4.1415 & 0.0113 & 0.0773 \\
		6   & 2.4620 & 3.7959 & 2.4608 & 3.7960 & 0.0488 & 0.0026 \\
		\bottomrule
	\end{tabular}
	\label{table-A}
\end{table}
\begin{figure}
	\setlength{\abovecaptionskip}{-0.05cm} 
	\setlength{\belowcaptionskip}{-0.3cm} 
	\subfigure[]{
		\begin{minipage}{0.5\linewidth}
			\centering
			\includegraphics[width=2.5in]{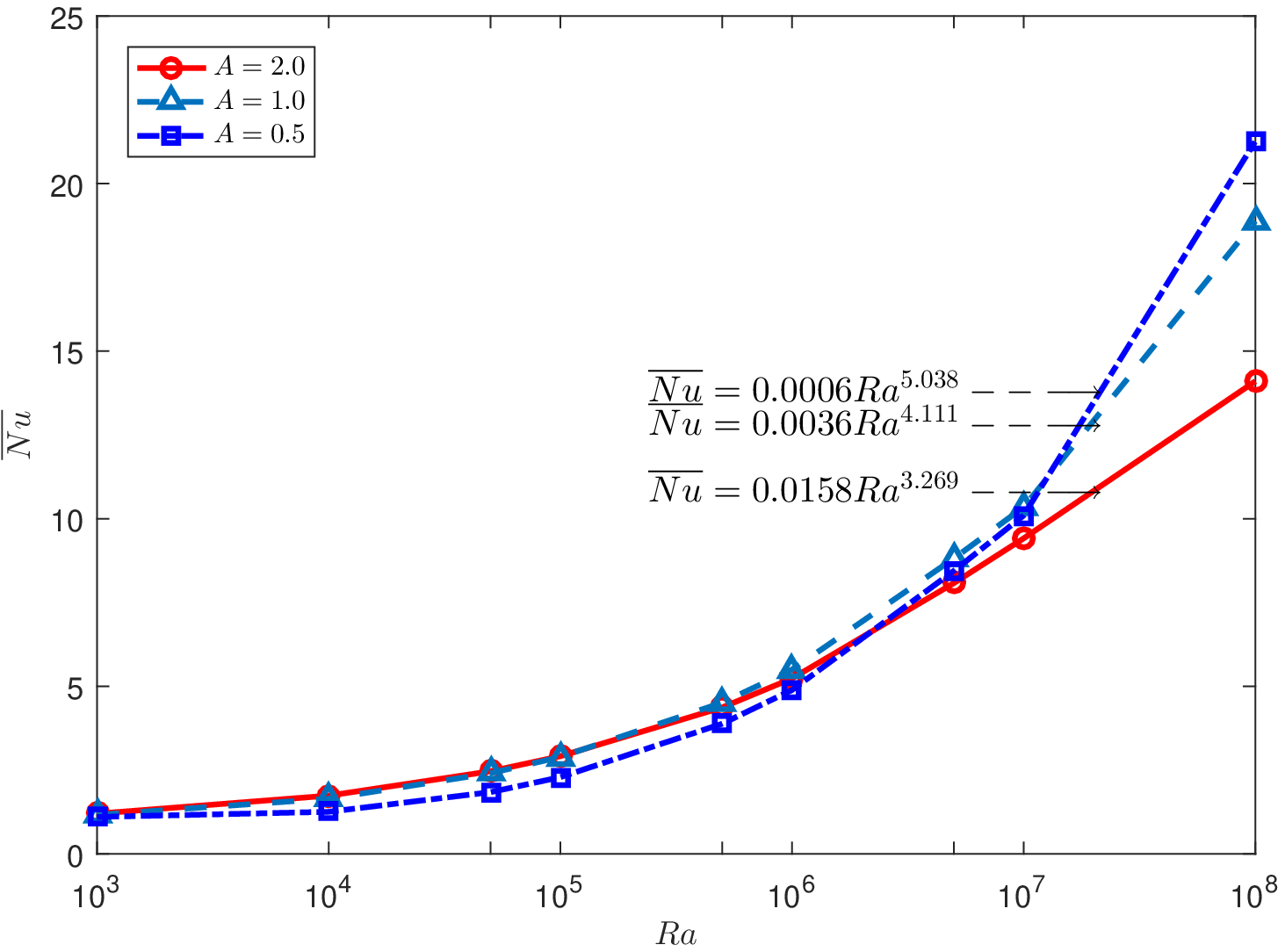}
			\label{fig-RaA-Nu}
	\end{minipage}}
	\subfigure[]{
		\begin{minipage}{0.5\linewidth}
			\centering
			\includegraphics[width=2.5in]{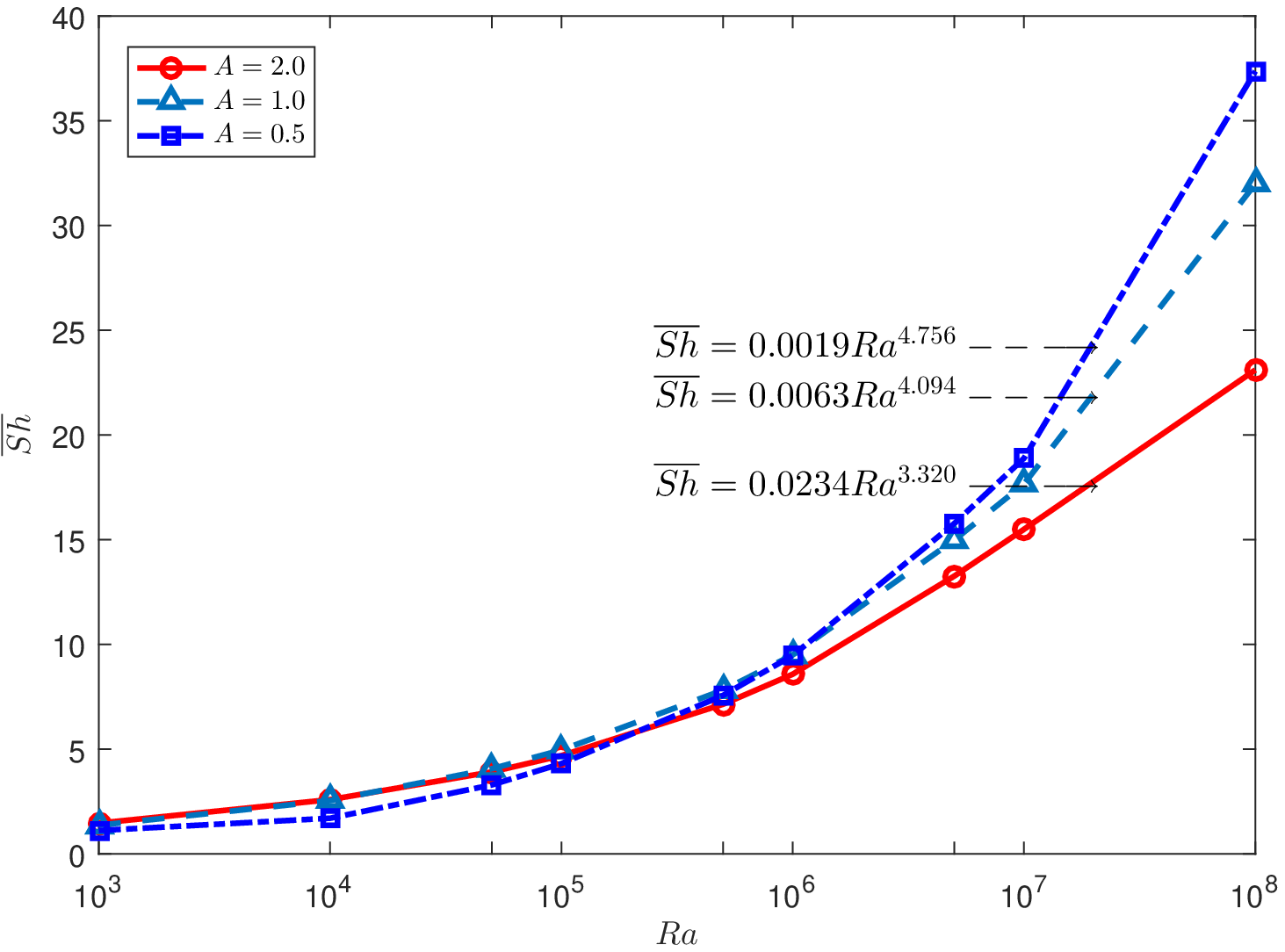}
			\label{fig-RaA-Sh}
	\end{minipage}}
	\caption{The average Nusselt number (a) and average Sherwood number (b) at different aspect ratios.}
	\label{fig-RaA}
\end{figure}
Moreover, through changing the Rayleigh number $Ra$, we can obtain scaling relations between average Nusselt and Sherwood numbers and Rayleigh number through fitting the data, as shown in Fig. \ref{fig-RaA}. As seen from this figure, when $A<1.0$, the increases of $\overline{Nu}$ and $\overline{Sh}$ are faster than the cases with $A>1.0$, which indicates the heat and mass transfer rates have the obvious changes in horizontal cavities. Here it should be noted that the values at $A=2.0$ and $Ra=10^8$ are time-averaged quantities.

\begin{figure}
	\setlength{\abovecaptionskip}{-0.05cm} 
	\setlength{\belowcaptionskip}{-0.3cm} 
	\subfigure[]{
		\begin{minipage}{0.5\linewidth}
			\centering
			\includegraphics[width=2.5in]{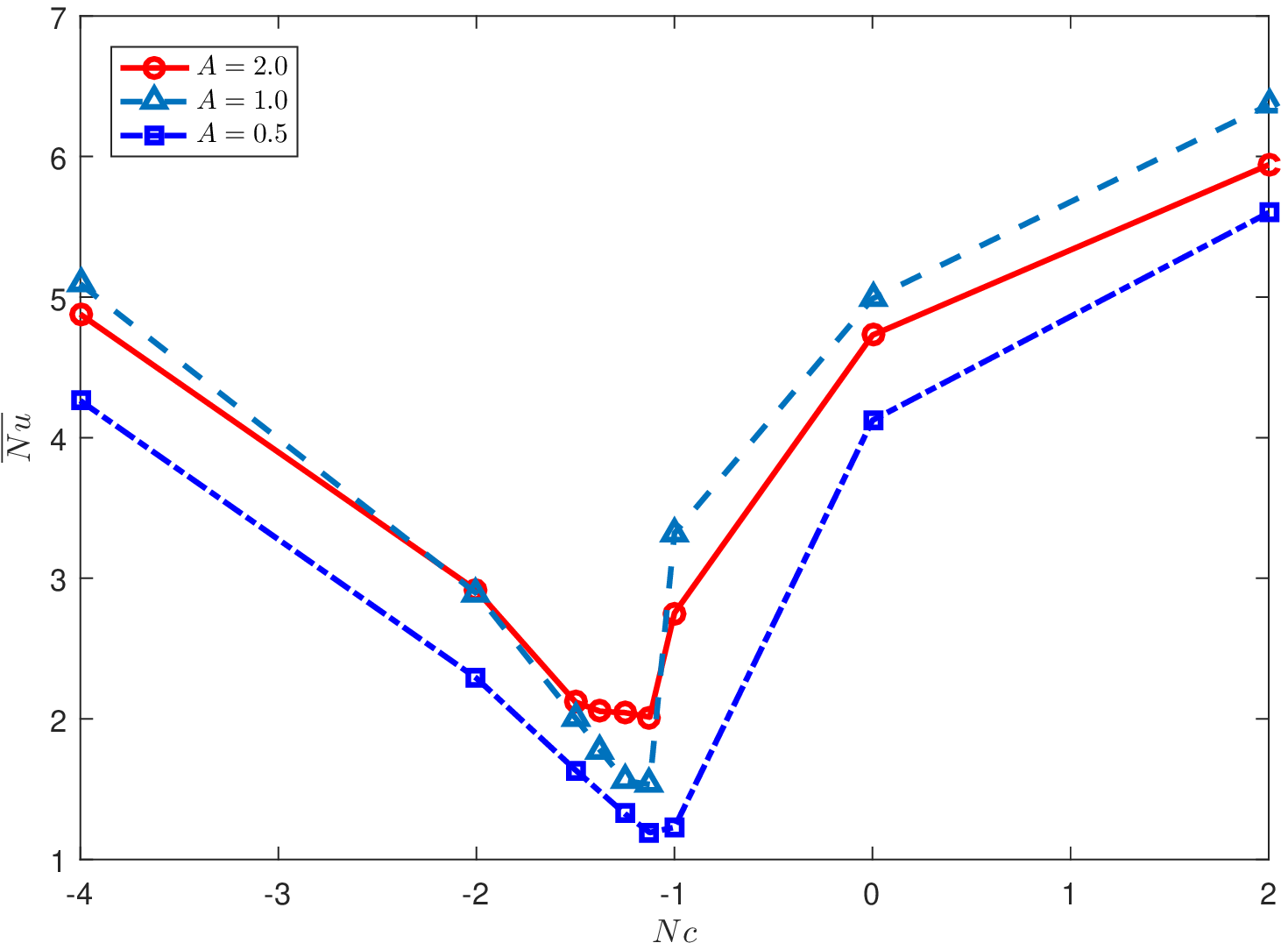}
			\label{fig-Nc-Nu}
	\end{minipage}}
	\subfigure[]{
		\begin{minipage}{0.5\linewidth}
			\centering
			\includegraphics[width=2.5in]{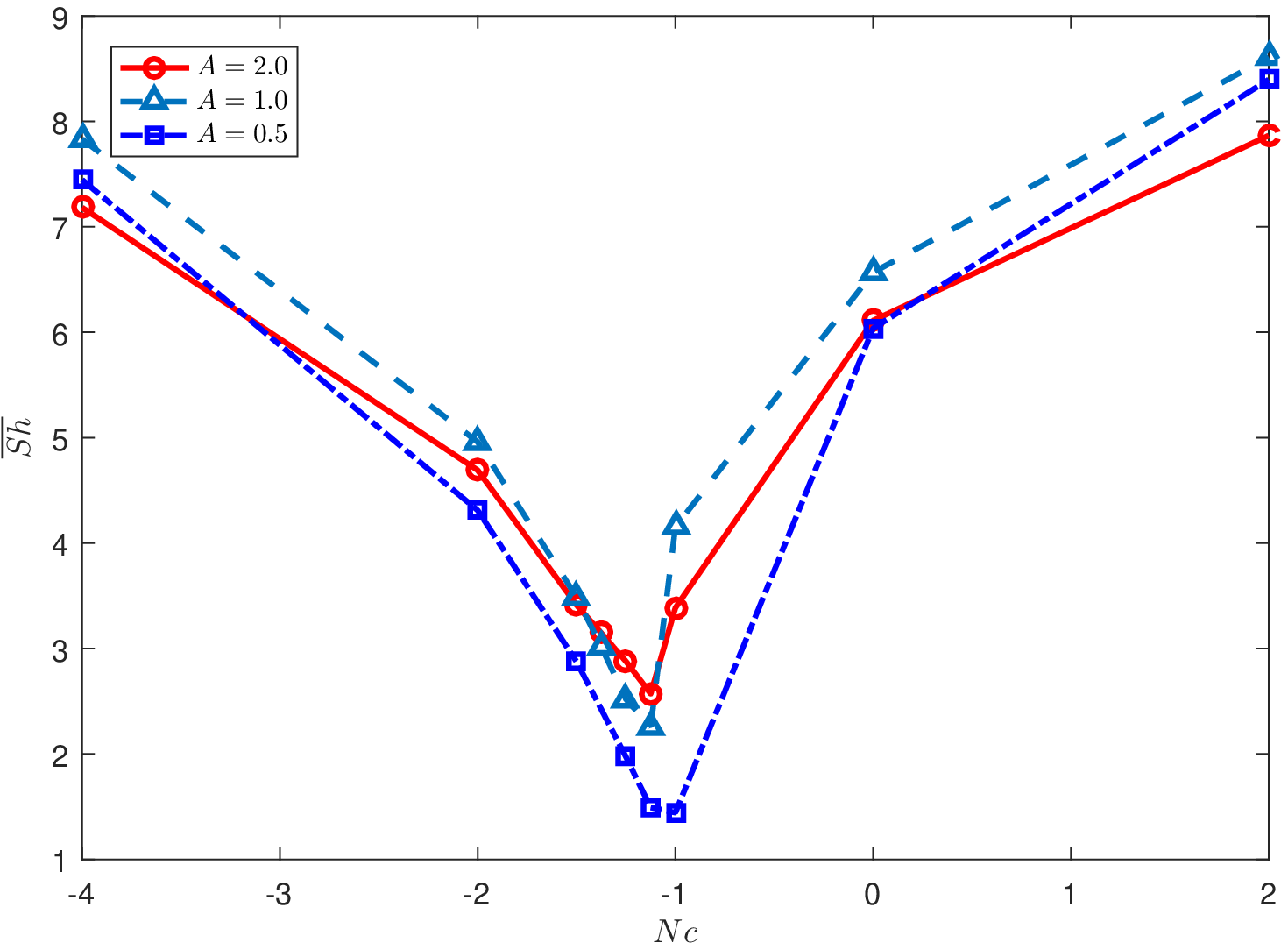}
			\label{fig-Nc-Sh}
	\end{minipage}}
	\caption{The average Nusselt number (a) and average Sherwood number (b) at different aspect ratios.}
	\label{fig-Nc}
\end{figure}

Next the influence of buoyancy ratio $Nc$ is also studied through fixing other parameters as $Ra=10^{5}$, $Pr=1.0$, $Le=2.0$, $S_{TC}=0.1$ and $D_{CT}=0.1$. One can see from Fig. \ref{fig-Nc} that the $\overline{Nu}$ and $\overline{Sh}$ at all aspect ratios are close to their minimum values at $Nc=-1.0$, where the thermal and solutal buoyancy forces are offset, and the results are more accurate for the horizontal cavities.
Here it is also worth noting that the values of average Nusselt and Sherwood numbers are time-averaged at $Nc=-1.0$ and $A=2.0$ due to the periodic flow.

\begin{figure}
	\setlength{\abovecaptionskip}{-0.05cm} 
	\setlength{\belowcaptionskip}{-0.3cm} 
	\subfigure[]{
		\begin{minipage}{0.5\linewidth}
			\centering
			\includegraphics[width=2.5in]{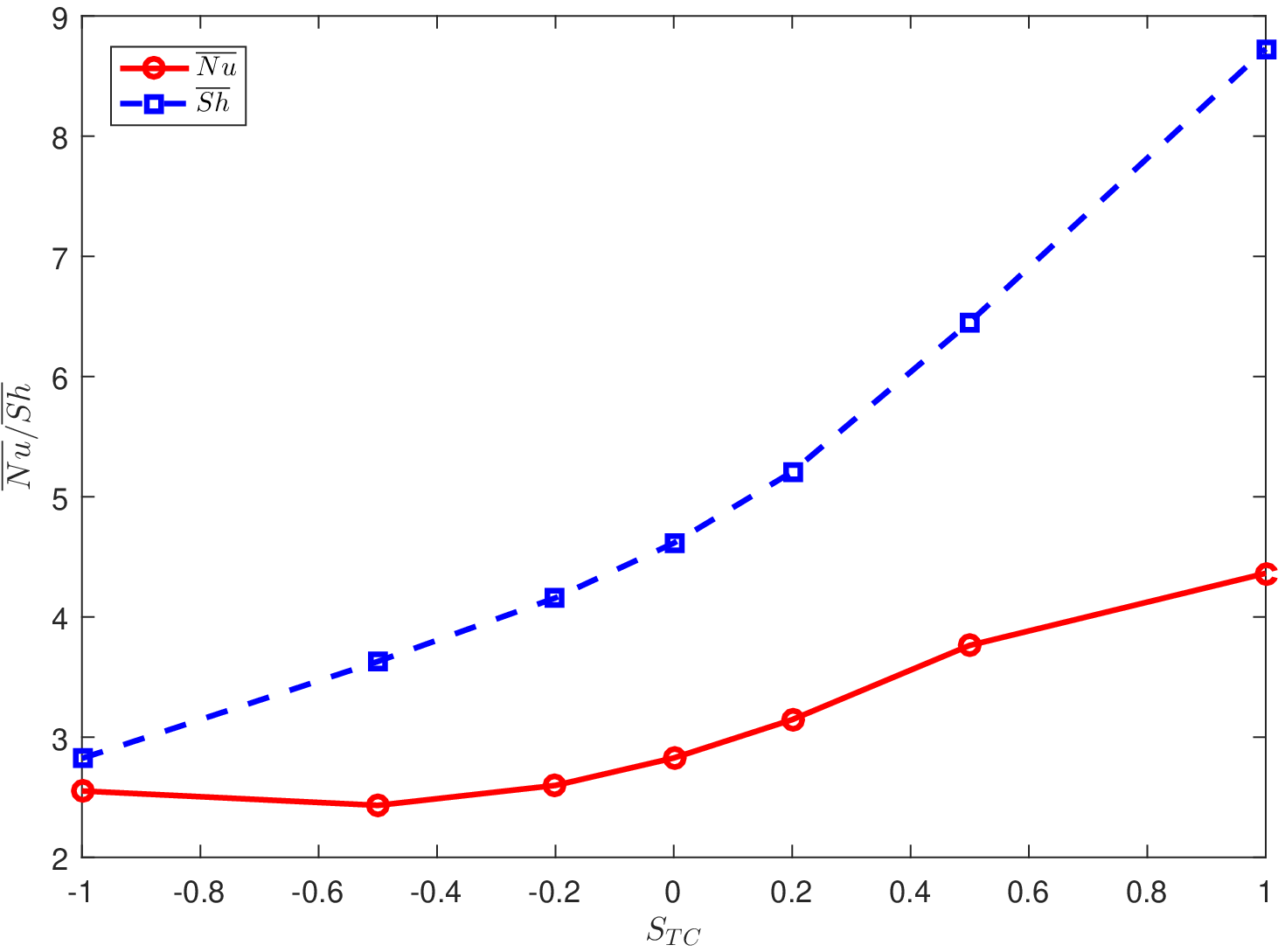}
			\label{fig-Stc}
	\end{minipage}}
	\subfigure[]{
		\begin{minipage}{0.5\linewidth}
			\centering
			\includegraphics[width=2.5in]{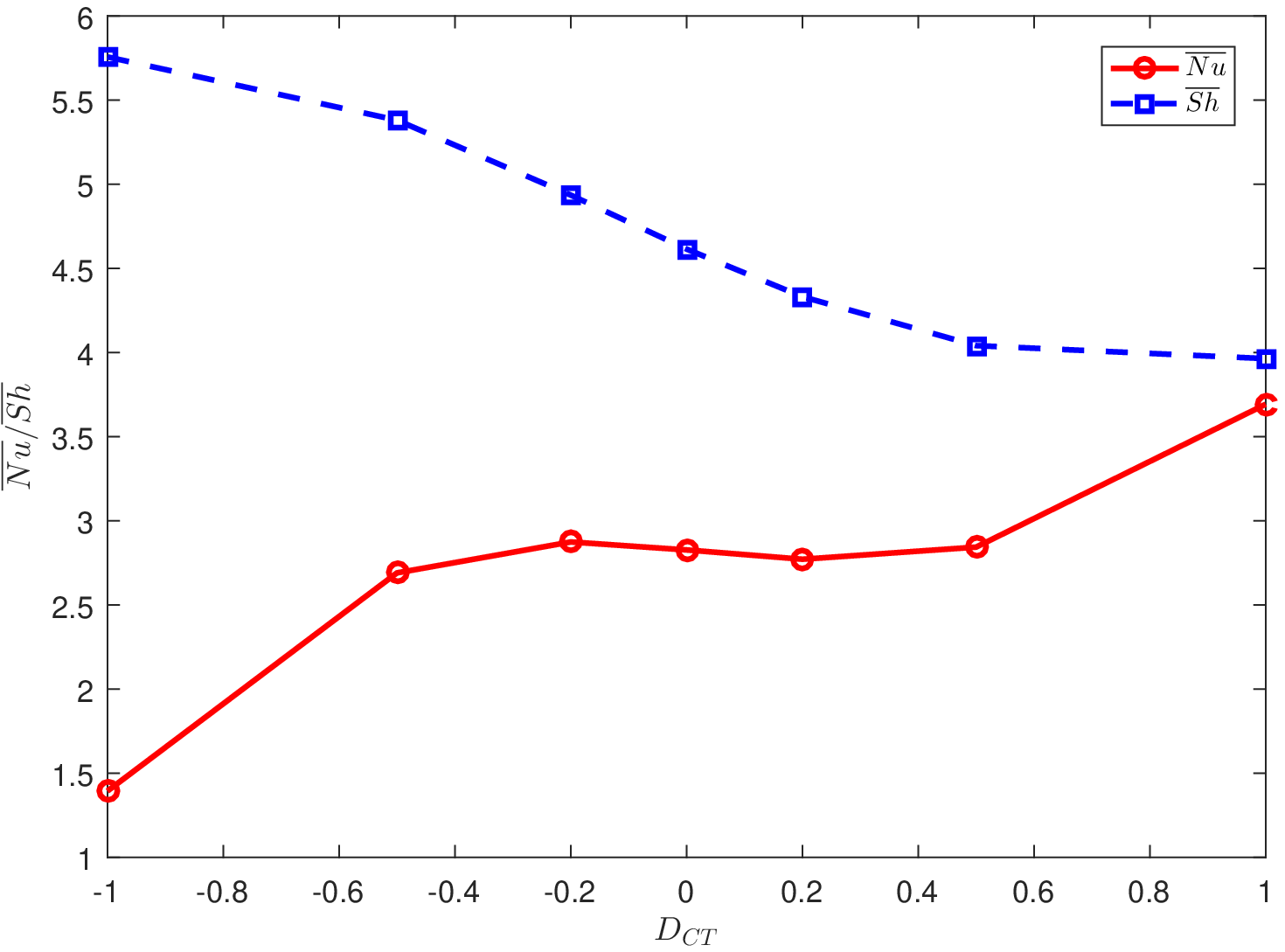}
			\label{fig-Dct}
	\end{minipage}}
	\caption{The average Nusselt and Sherwood numbers at different Soret and Dufour factors, [(a) Soret factor $S_{TC}$ at $D_{CT}=0$ and (b) Dufour factor $D_{CT}$ at $S_{TC}=0$].}
	\label{fig-DS}
\end{figure}

Finally, we focus on the effects of the Soret and Dufour factors at $Ra=10^{5}$, $Pr=1.0$, $Le=2.0$, $Nc=-2.0$ and $A=2.0$. When Soret and Dufour factors are equal to each other, $\overline{Nu}$ and $\overline{Sh}$ increase with the increase of Soret and Dufour factors, as presented in Table \ref{table-DS} where the relative errors are within 0.2\%, compared with \cite{Ren2016IJHMT}.  
In addition, we take $D_{CT}=0$, and present the effect of Soret factor in Fig. \ref{fig-Stc}. From this figure one can see that the average Sherwood number increases in Soret factor, while the average Nusselt number first decreases when $S_{TC}$ is less than $-0.5$, and then increases in Soret factor when $S_{TC}$ is larger than $-0.5$. However, as shown in Fig. \ref{fig-Dct}, when we fix $S_{TC}=0$ and increase the Dufour factor, the average Sherwood number decreases, while the average Nusselt number increases in a large range of $D_{CT}$ but decreases in a small region near $D_{CT}=0$.
\begin{table}
	\centering
	\caption{The average Nusselt and Sherwood numbers of the DDC system at $Ra=10^{5}$, $Pr=1.0$, $Le=2.0$, $Nc=-2.0$, and $A=2.0$.}
	\begin{tabular}{cccccc}
		\toprule
		& & \multicolumn{2}{c}{Present} & \multicolumn{2}{c}{Ref. \cite{Ren2016IJHMT}} \\
		\cline{3-6}
		Soret factor $S_{TC}$ & Dufour factor $D_{CT}$ & $\overline{Nu}$ & $\overline{Sh}$ & $\overline{Nu}$ & $\overline{Sh}$ \\
		\midrule
		-0.2 & -0.2 & 2.5609 & 4.3020 &   -    &   -    \\
		-0.1 & -0.1 & 2.7079 & 4.4842 &   -    &   -    \\
		0    & 0    & 2.8265 & 4.6144 & 2.8275 & 4.6232 \\
		0.05 & 0.05 & 2.8747 & 4.6597 & 2.8753 & 4.6681 \\
		0.1  & 0.1  & 2.9151 & 4.6914 & 2.9153 & 4.6993 \\
		0.15 & 0.15 & 2.9475 & 4.7091 & 2.9473 & 4.7164 \\
		0.2  & 0.2  & 2.9715 & 4.7123 & 2.9709 & 4.7191 \\
		\bottomrule
	\end{tabular}
	\label{table-DS}
\end{table}

\section{Conclusions}
\label{Conclusions}
In this paper, a LB model is proposed for the general CDF system in $d$ dimensional space. Through the direct Taylor expansion, the governing equations of the general system can be recovered correctly from the proposed LB model. In the present LB model, the cross-diffusion terms are modeled through introducing some extra collision operators, and the proper auxiliary source terms are also constructed, which can be used to avoid the discretization of some derivative terms. Additionally, the computational scheme of pressure actually contains the external force but can be simplified under the incompressible condition. We validate the present LB model by two important CDF systems, i.e., CF system and DDC system with Soret and Dufour effects, and find the results are in good agreement with the analytical solutions and the previous works. Moreover, in the DDC system, the influence of Rayleigh number at different aspect ratios is considered, and the power-law relations between the average Nusselt and Sherwood numbers and the Rayleigh number are observed, which indicate that the heat and mass transfer rates increase faster with the increase of the Rayleigh number in horizontal cavities.
In addition, the average Nusselt and Sherwood numbers are close to their minimum values when buoyancy ratio approaches to $-1$, which is more accurate for the horizontal cavities.
Finally, the effects of Soret and Dufour factors are also investigated, and the results show that the average Nusselt number increases in a large interval of Soret and Dufour factors, while the average Sherwood number increases with Soret factor but decreases with the Dufour factor.

\section*{Acknowledgments}
This work is supported by the National Natural Science Foundation of China (Grant No. 51836003), and the National Key Research and Development Program of China (Grant No. 2017YFE0100100).

\appendix
\section{The lattice Boltzmann model for incompressible Navier-Stokes equations}\label{appendix}
In this appendix, we will adopt the direct Taylor expansion method to recover the incompressible Navier-Stokes equations from the present LB model.
We first apply the Taylor expansion to Eq.\,(\ref{evolution3}), similar to Eq.\,(\ref{taylorEX1}), and can obtain the following equations at different orders of $\Delta t$,
\begin{subequations}
	\begin{equation}\label{order1NS}
	D_{i}h_{i}^{eq}=-\frac{\omega}{\Delta t}h_{i}^{ne}+(1-\frac{\omega}{2})F_{i}+O(\Delta t),
	\end{equation}
	\begin{equation}\label{order2NS}
	D_{i}(h_{i}^{eq}+h_{i}^{ne})+\frac{\Delta t}{2}D_{i}^{2}h_{i}^{eq}=-\frac{\omega}{\Delta t}h_{i}^{ne}+(1-\frac{\omega}{2})F_{i}+O(\Delta t^2),
	\end{equation}
\end{subequations}
where $h_{i}^{ne}$ is the non-equilibrium part of $h_{i}$. According to Eq.\,(\ref{order1NS}), we have
\begin{equation}\label{D-order1NS}
\frac{\Delta t}{2}D_{i}^{2}h_{i}^{eq}=-\frac{1}{2}D_{i}\omega h_{i}^{ne}+\frac{\Delta t}{2}D_{i}(1-\frac{\omega}{2})F_{i}+O(\Delta t^2).
\end{equation}
Then substituting Eq.\,(\ref{D-order1NS}) into Eq.\,(\ref{order2NS}), one can obtain
\begin{equation}\label{order2-NS}
\begin{aligned}
D_{i}h_{i}^{eq}+D_{i}(1-\frac{\omega}{2})h_{i}^{ne}+\frac{\Delta t}{2}D_{i}(1-\frac{\omega}{2})F_{i}=-\frac{\omega}{\Delta t}h_{i}^{ne}+(1-\frac{\omega}{2})F_{i}+O(\Delta t^2).
\end{aligned}
\end{equation}
To give the correct Navier-Stokes equations, the distribution functions $h_{i}$, $h_{i}^{eq}$ and $F_{i}$ satisfy the following conditions,
\begin{subequations}\label{momentNS}
	\begin{equation}\label{momentNSh}
	\sum_{i} h_{i}=\sum_{i}h_{i}^{eq}=\rho_{0},\ \sum_{i} \mathbf{c}_{i} h_{i}^{eq}=\mathbf{u},\
	 \sum_{i}\mathbf{c}_{i}\mathbf{c}_{i} h_{i}^{eq}=p \mathbf{I}+\mathbf{uu},\ \sum_{i}\mathbf{c}_{i}\mathbf{c}_{i}\mathbf{c}_{i}h_{i}^{eq}=\Delta\cdot\mathbf{u},
	\end{equation}
	\begin{equation}
	\sum_{i} F_{i}=0,\ \sum_{i} \mathbf{c}_{i} F_{i}=\mathbf{F},\ \sum_{i} \mathbf{c}_{i} \mathbf{c}_{i} F_{i}=\varphi(\mathbf{uF}+\mathbf{Fu}),
	\end{equation}
\end{subequations}
where $\Delta$ is a fourth-order tensor giving by $\delta_{\alpha\beta}\delta_{\theta\gamma}+\delta_{\beta\theta}\delta_{\alpha\gamma}+\delta_{\alpha\theta}\delta_{\beta\gamma}$. From Eqs.\,(\ref{momentNSh}) and (\ref{sum-u}), we can get
\begin{equation}
\sum_{i}\mathbf{c}_{i}h_{i}^{ne}=\sum_{i}\mathbf{c}_{i}h_{i}-\sum_{i}\mathbf{c}_{i}h_{i}^{eq}=-\frac{\Delta t}{2}\mathbf{F}.
\end{equation}
Using above relations, we can derive the zeroth and first order moments of Eqs.\,(\ref{order1NS}) and (\ref{order2-NS}) at the orders of $O(\Delta t)$ and $O(\Delta t^2)$,
\begin{subequations}\label{orderdt1}
	\begin{equation}\label{order11NS}
	\nabla\cdot\mathbf{u}=O(\Delta t),
	\end{equation}
	\begin{equation}\label{order12NS}
	\partial_{t}\mathbf{u}+\nabla\cdot (p\mathbf{I}+\mathbf{uu})=\mathbf{F}+O(\Delta t),
	\end{equation}
\end{subequations}
\begin{subequations}\label{orderdt2}
	\begin{equation}\label{order21NS}
	\nabla\cdot\mathbf{u}=O(\Delta t^2),
	\end{equation}
	\begin{equation}\label{order22NS}
	\partial_{t}\mathbf{u}+\nabla\cdot(p\mathbf{I}+\mathbf{uu})+\nabla\cdot (1-\frac{\omega}{2})\left[\sum_{i} \mathbf{c}_{i}\mathbf{c}_{i} h_{i}^{ne}+\frac{\Delta t}{2} \varphi(\mathbf{u}\mathbf{F}+\mathbf{F}\mathbf{u})\right]=\mathbf{F}+O(\Delta t^2),
	\end{equation}
\end{subequations}	
where the term $\sum_{i}\mathbf{c}_{i}\mathbf{c}_{i}h_{i}^{ne}$ can be evaluated by Eq.\,(\ref{order1NS}),
\begin{equation}\label{cicihine}
	\begin{aligned}
	\sum_{i}\mathbf{c}_{i}\mathbf{c}_{i}h_{i}^{ne}=&-\frac{\Delta t}{\omega}\sum_{i} \mathbf{c}_{i} \mathbf{c}_{i} \left[ D_{i}h_{i}^{eq}-(1-\frac{\omega}{2})F_{i}\right]+O(\Delta t^2)\\
	=& -\frac{\Delta t}{\omega}\left[\partial_{t}\sum_{i}\mathbf{c}_{i}\mathbf{c}_{i}h_{i}^{eq}+\nabla\cdot\sum_{i}\mathbf{c}_{i}\mathbf{c}_{i}\mathbf{c}_{i}h_{i}^{eq}-(1-\frac{\omega}{2})\sum_{i}\mathbf{c}_{i}\mathbf{c}_{i}F_{i}\right]+O(\Delta t^2)\\
	=&-\frac{\Delta t}{\omega}\left[\partial_{t}(p\mathbf{I}+\mathbf{uu})+\nabla\cdot(\Delta\cdot\mathbf{u})-(1-\frac{\omega}{2})\varphi(\mathbf{uF}+\mathbf{Fu})\right]+O(\Delta t^2)\\
	=&-\frac{\Delta t}{\omega}\left[\partial_{t}(p\mathbf{I}+\mathbf{uu})+c_{s}^{2}[\nabla\mathbf{u}+(\nabla\mathbf{u})^{T}]-(1-\frac{\omega}{2})\varphi(\mathbf{uF}+\mathbf{Fu})\right]+O(\Delta t^2).
	\end{aligned}
\end{equation}
Based on Eq.\,(\ref{orderdt1}), we get
\begin{equation}\label{ptuu}
\begin{aligned}
\partial_{t}(u_{\alpha}u_{\beta})=&u_{\alpha}(F_{\beta}-\nabla_{\beta}p-\nabla_{\gamma}u_{\beta}u_{\gamma})+(F_{\alpha}-\nabla_{\alpha}p-\nabla_{\gamma}u_{\alpha}u_{\gamma})u_{\beta}+O(Ma\Delta t)\\
=&u_{\alpha}F_{\beta}+F_{\alpha}u_{\beta}-u_{\alpha}\nabla_{\beta}p-\nabla_{\alpha}pu_{\beta}-\nabla_{\gamma}u_{\alpha}u_{\beta}u_{\gamma}+O(Ma\Delta t)\\
=&u_{\alpha}F_{\beta}+F_{\alpha}u_{\beta}+O(Ma\Delta t+Ma^3),
\end{aligned}
\end{equation}
where $Ma$ is the Mach number. Substituting Eq.\,(\ref{ptuu}) into Eq.\,(\ref{cicihine}) and with the help of $\partial_{t}p=O(Ma^2)$, one can obtain
\begin{align}\label{cicihine-1}
\begin{autobreak}
\sum_{i}\mathbf{c}_{i}\mathbf{c}_{i}h_{i}^{ne}+\frac{\Delta t}{2} \varphi(\mathbf{u}\mathbf{F}+\mathbf{F}\mathbf{u})=
\Delta t\frac{\varphi-1}{\omega}(\mathbf{u}\mathbf{F}+\mathbf{F}\mathbf{u})
-\frac{c_{s}^{2}\Delta t}{\omega}[\nabla\mathbf{u}+(\nabla\mathbf{u})^{T}]
+O(Ma^2\Delta t +\Delta t^2).
\end{autobreak}
\end{align}
According to the value of $\varphi$, we have the following two cases.

\noindent\textbf{Case 1:} $\varphi=0$. Due to the fact $\mathbf{u}\mathbf{F}=O(Ma^2)$, the first term on the right hand side of Eq.\,(\ref{cicihine-1}) can be absorbed into the truncation term $O(Ma^2\Delta t+\Delta t^2)$. Actually, the derivation process of Eq.\,(\ref{ptuu}) is not necessary since $\partial_{t}\mathbf{uu}=O(Ma^2)$. In this case, the force term $F_{i}$ can be simplified.

\noindent\textbf{Case 2:} $\varphi=1$. Under this condition, the first term on the right hand side of Eq.\,(\ref{cicihine-1}) is zero.

In both cases, if we substitute Eq.\,(\ref{cicihine-1}) into Eq.\,(\ref{order22NS}), and omit the truncation terms $O(\Delta t^2)$ in Eq.\,(\ref{order21NS}) and $O(Ma^2\Delta t+\Delta t^2)$ in Eq.\,(\ref{order22NS}), one can get the incompressible Navier-Stokes equations:
\begin{subequations}
	\begin{equation}
	\nabla\cdot\mathbf{u}=0,
	\end{equation}
	\begin{equation}
	\mathbf{u}_{t}+\nabla\cdot\mathbf{uu}=-\nabla p +\nabla\cdot\nu [\nabla \mathbf{u}+(\nabla\mathbf{u})^{T}]+\mathbf{F},
	\end{equation}	
\end{subequations}
where $\nu=(\frac{1}{\omega}-\frac{1}{2})c_{s}^{2}\Delta t$.

Now let focus on the computation of pressure. From Eq.\,(\ref{order1NS}) one can obtain 
\begin{equation}\label{h-neq}
h_{i}^{ne}=-\frac{\Delta t}{\omega}\left[D_{i}h_{i}^{eq}-(1-\frac{\omega}{2})F_{i}\right]+O(\Delta t^2).
\end{equation}
Taking the zeroth-direction of Eqs.\,(\ref{heq}) and (\ref{h-neq}), we have
\begin{equation}\label{heq0}
h_{0}^{eq}=\frac{(W_{0}-1)p}{c_{s}^{2}}+\rho_{0}-W_{0}\frac{\mathbf{u}\cdot\mathbf{u}}{2c_{s}^{2}},
\end{equation}
\begin{equation}\label{hne0}
h_{0}^{ne}=-\frac{\Delta t}{\omega}\left[\partial_{t}h_{0}^{eq}-(1-\frac{\omega}{2})F_{0}\right]+O(\Delta t^2).
\end{equation}
Owing to the fact that $\partial_{t}h_{0}^{eq}$ is order of $O(Ma^2)$, Eq.\,(\ref{hne0}) can be simplified by
\begin{equation}\label{hne0-1}
h_{0}^{ne}=\frac{\Delta t}{\omega}(1-\frac{\omega}{2})F_{0}+O(Ma^2\Delta t+\Delta t^2).
\end{equation}
According to Eqs.\,(\ref{heq0}) and (\ref{hne0-1}), and based on $h_{i}=h_{i}^{eq}+h_{i}^{ne}$, we have
\begin{equation}
\begin{aligned}
\frac{(1-W_{0})p}{c_{s}^{2}}=&\rho_{0}-(h_{0}-h_{0}^{ne})-W_{0}\frac{\mathbf{u}\cdot\mathbf{u}}{2c_{s}^{2}}\\
=&\rho_{0}-\left[\sum_{i}h_{i}-\sum_{i\neq 0}h_{i}\right]-W_{0}\frac{\mathbf{u}\cdot\mathbf{u}}{2c_{s}^{2}}+\frac{\Delta t}{\omega}(1-\frac{\omega}{2})F_{0}+O(Ma^2\Delta t+\Delta t^2)\\
=&\sum_{i\neq 0}h_{i}-W_{0}\frac{\mathbf{u}\cdot\mathbf{u}}{2c_{s}^{2}}+\frac{\Delta t}{\omega}(1-\frac{\omega}{2})F_{0}+O(Ma^2\Delta t+\Delta t^2).
\end{aligned}
\end{equation}
Ignoring the truncation error terms of $O(Ma^2\Delta t+\Delta t^2)$, we can obtain the computational scheme for pressure,
\begin{equation}
\begin{aligned}
p=&\frac{c_{s}^{2}}{1-W_{0}}\left[\sum_{i\neq 0}h_{i}-W_{0}\frac{\mathbf{u}\cdot\mathbf{u}}{2c_{s}^{2}}+\frac{\Delta t}{\omega}(1-\frac{\omega}{2})F_{0}\right]\\
=&\frac{c_{s}^{2}}{1-W_{0}}\left[\sum_{i\neq 0}h_{i}-W_{0}\frac{\mathbf{u}\cdot\mathbf{u}}{2c_{s}^{2}}-\varphi\Delta t(\frac{1}{\omega}-\frac{1}{2})\frac{\mathbf{u}\cdot\mathbf{F}}{c_{s}^{2}}\right].
\end{aligned}
\end{equation}
\bibliographystyle{elsarticle-num} 
\bibliography{reference}

\begin{thebibliography}{10}
\expandafter\ifx\csname url\endcsname\relax
  \def\url#1{\texttt{#1}}\fi
\expandafter\ifx\csname urlprefix\endcsname\relax\def\urlprefix{URL }\fi
\expandafter\ifx\csname href\endcsname\relax
  \def\href#1#2{#2} \def\path#1{#1}\fi

\bibitem{KS1970JTB}
E.~F. Keller, L.~A. Segel, Initiation of slime mold aggregation viewed as an
  instability, Journal of Theoretical Biology 26 (1970) 399--415.

\bibitem{KS1971JTB}
E.~F. Keller, L.~A. Segel, Model for chemotaxis, Journal of Theoretical Biology
  30 (1971) 225--234.

\bibitem{Shige1979JTB}
N.~Shigesada, K.~Kawasaki, E.~Teramoto, Spatial segregation of interacting
  species, Journal of Theoretical Biology 79~(1) (1979) 83--99.

\bibitem{Hitt2017MMMAS}
S.~Hittmeir, H.~Ranetbauer, C.~Schmeiser, M.-T. Wolfram, Derivation and
  analysis of continuum models for crossing pedestrian traffic, Mathematical
  Models and Methods in Applied Sciences 27~(07) (2017) 1301--1325.

\bibitem{Curtiss1999IECR}
C.~F. Curtiss, R.~B. Bird, Multicomponent diffusion, Industrial \& Engineering
  Chemistry Research 38~(7) (1999) 2515--2522.

\bibitem{Hillesdon1995BMB}
A.~J. Hillesdon, T.~J. Pedley, J.~O. Kessler, Bioconvection in suspensions of
  oxytactic beacteria: linear theory, Bulletin of Mathematical Biology 57
  (1995) 299--303.

\bibitem{Dombrowski2004PRL}
C.~Dombrowski, L.~Cisneros, S.~Chatkaew, R.~E. Goldstein, J.~O. Kessler,
  Self-concentration and large-scale coherence in bacterial dynamics, Physical
  Review Letters 93 (2004) 098103.

\bibitem{Tuval2005PNAS}
I.~Tuval, L.~Cisneros, C.~Dombrowski, C.~W. Wolgemuth, J.~O. Kessler, R.~E.
  Goldstein, Bacterial swimming and oxygen transport near contact lines,
  Proceeding of the National Academy of Sciences of the United States of
  America 102~(7) (2005) 2277--2282.

\bibitem{Trevisan1987JHT}
O.~V. Trevisan, A.~Bejan, Combined heat and mass transfer by natural convection
  in a vertical enclosure, Journal of Heat Transfer 109~(1) (1987) 104--112.

\bibitem{Gaikwad2007IJNLM}
S.~N. Gaikwad, M.~S. Malashetty, K.~R. Prasad, An analytical study of linear
  and non-linear double diffusive convection with {Soret} and {Dufour} effects
  in couple stress fluid, International Journal of Non-Linear Mechanics 42~(7)
  (2007) 903--913.

\bibitem{Budroni2015PRE}
M.~A. Budroni, Cross-diffusion-driven hydrodynamic instabilities in a
  double-layer system: General classification and nonlinear simulations,
  Physical Review E 92~(6) (2015) 063007.

\bibitem{Bend2018MMAS}
M.~Bendahmane, F.~Karami, M.~Zagour, Kinetic‐fluid derivation and
  mathematical analysis of the cross‐diffusion–{Brinkman} system,
  Mathematical Methods in the Applied Sciences 41~(16)  6288--6311.

\bibitem{KIM2019IJHMT}
M.~C. Kim, K.~H. Song, Theoretical and numerical analyses of the effect of
  cross-diffusion on the gravitational instability in ternary mixtures,
  International Journal of Heat and Mass Transfer 43 (2019) 118511.

\bibitem{Rag2019ZAMP}
K.~R. Raghunatha, I.~S. Shivakumara, M.~S. Swamy, Effect of cross-diffusion on
  the stability of a triple-diffusive {Oldroyd-B} fluid layer, Zeitschrift für
  angewandte Mathematik und Physik 70~(100) (2019).

\bibitem{Atlas2020AMM}
A.~Atlas, M.~Bendahmane, F.~Karami, D.~Meskine, Kinetic-fluid derivation and
  mathematical analysis of anonlocal cross-diffusion–fluid system, Applied
  Mathematical Modelling 82 (2020) 379--408.

\bibitem{Chertock2012JFM}
A.~Chertock, K.~Fellner, A.~Kurganov, A.~Lorz, P.~A. Markowich, Sinking,
  merging and stationary plumes in a coupled chemotaxis-fluid model: a
  high-resolution numerical approach, Journal of Fluid Mechanics 694 (2012)
  155--190.

\bibitem{Sheu2014CF}
T.~W. Sheu, C.~Y. Chiang, Numerical investigation of chemotaxic phenomenon in
  incompressible viscous fluid flow, Computer and Fluids 103 (2014) 290--306.

\bibitem{Lee2015EJMBF}
H.~G. Lee, J.~Kim, Numerical investigation of falling bacterial plumes caused
  by bioconvection in a three-dimensional chamber, European Journal of
  Mechanics B/Fluids 52 (2015) 120--130.

\bibitem{Deleuze2016CF}
Y.~Deleuze, C.-Y. Chiang, M.~Thiriet, T.~W. Sheu, Numerical study of plume
  patterns in a chemotaxis-diffusion-convection coupling system, Computer and
  Fluids 126 (2016) 58--70.

\bibitem{Nithyadevi2009IJHFF}
N.~Nithyadevi, R.-J. Yang, Double diffusive natural convection in a partially
  heated enclosure with {Soret} and {Dufour} effects, International Journal of
  Heat and Fluid Flow 30 (2009) 902--910.

\bibitem{Beg2011IJHMT}
O.~A. B\'{e}g, V.~R. Prasad, B.~Vasu, N.~B. Reddy, Q.~Li, R.~Bhargava, Free
  convection heat and mass transfer from an isothermal sphere to a micropolar
  regime with {Soret/Dufour} effects, International Journal of Heat and Mass
  Transfer 54 (2011) 9--18.

\bibitem{Cheng2012ICHMT}
C.-Y. Cheng, Soret and {Dufour} effects on free convection heat and mass
  transfer from an arbitrarily inclined plate in a porous medium with constant
  wall temperature and concentration, International Communications in Heat and
  Mass Transfer 39 (2012) 72--77.

\bibitem{Wang2014IJHMT}
J.~Wang, M.~Yang, Y.~Zhang, Onset of double-diffusive convection in horizontal
  cavity with {Soret} and {Dufour} effects, International Journal of Heat and
  Mass Transfer 78 (2014) 1023--1031.

\bibitem{Wang2016IJTS}
J.~Wang, M.~Yang, Y.-L. He, Y.~Zhang, Oscillatory double-diffusive convection
  in a horizontal cavity with {Soret} and {Dufour} effects, International
  Journal of Thermal Sciences 106 (2016) 57--69.

\bibitem{Higuera1989EPL}
F.~J. Higuera, S.~Succi, R.~Benzi, Lattice gas dynamics with enhanced
  collisions, Europhysics Letters 9~(4) (1989) 345--349.

\bibitem{Benzi1992PR}
R.~Benzi, S.~Succi, M.~Vergassloa, The lattice {Boltzmann} equation: theory and
  applications, Physics Reports 222 (1992) 145--197.

\bibitem{Qian1995ARCP}
Y.~Qian, S.~Succi, S.~A. Orszag, Recent advances in lattice {Boltzmann}
  computing, Annual Reviews of Computational Physics 3 (1995) 195--242.

\bibitem{Chen1998ARFM}
S.~Chen, G.~D. Doolen, Lattice {Boltzmann} method for fluid flows, Annual
  Reviews of Fluid Mechanics 30 (1998) 329--364.

\bibitem{Aidun2010ARFM}
C.~K. Aidun, J.~R. Clausen, Lattice-{Boltzmann} method for complex flows,
  Annual Review of Fluid Mechanics 42~(1) (2010) 439--472.

\bibitem{Wang2019Capillarity}
H.~Wang, X.~Yuan, H.~Liang, Z.~Chai, B.~Shi, A brief review of the
  phase-field-based lattice {Boltzmann} method for multiphase flows,
  Capillarity 2~(2) (2019) 33--52.

\bibitem{Dawson1993JCP}
S.~P. Dawson, S.~Chen, G.~D. Doolen, Lattice {Boltzmann} computations for
  reaction-diffusion equations, The Journal of Chemical Physics 98~(2) (1993)
  1514.

\bibitem{Blaak2000CPC}
R.~Blaak, P.~M. Sloot, Lattice dependence of reaction-diffusion in lattice
  {Boltzmann} modeling, Computer Physics Communications 129 (2000) 256--266.

\bibitem{He2000MS}
X.~He, N.~Li, B.~Goldstein, Lattice {Boltzmann} simulation of
  diffusion-convection systems with surface chemical reaction, Molecular
  Simulation 25~(3) (2000) 145--156.

\bibitem{Shi2009PRE}
B.~Shi, Z.~Guo, Lattice {Boltzmann} model for nonlinear convection-diffusion
  equation, Physical Review E 79 (2009) 016701.

\bibitem{Chai2013PRE}
Z.~Chai, T.~S. Zhao, Lattice {Boltzmann} model for the convection-diffusion
  equation, Physical Review E 87 (2013) 063309.

\bibitem{Succi2001}
S.~Succi, The Lattice Boltzmann Equation for Fluid Dynamics and Beyond, Oxford
  University Press, 2001.

\bibitem{Kruger2017}
T.~Krüger, H.~Kusumaatmaja, A.~Kuzmin, O.~Shardt, G.~Silva, E.~M. Viggen, The
  Lattice {Boltzmann} Method: Principles and Practice, Oxford University Press,
  2017.

\bibitem{Hilpert2005JMB}
M.~Hilpert, Lattice {Boltzmann} model for bacteria chemotaxis, Journal of
  Mathematical Biology 51 (2005) 302--332.

\bibitem{Yu2007ICCS}
X.~Yu, Z.~Guo, B.~Shi, Numerical study of cross diffusion effects on double
  diffusion convection with lattice {Boltzmann} method, International
  Conference on Computational Science 4487 (2007) 810--817.

\bibitem{Yang2014CMA}
X.~Yang, B.~Shi, Z.~Chai, Coupled lattice {Boltzmann} method for generalized
  {Keller-Segel} chemotaxis model, Computers and Mathematics with Applications
  68 (2014) 1653--1670.

\bibitem{Chai2019SIAM}
Z.~Chai, H.~Liang, R.~Du, B.~Shi, A lattice {Boltzmann} model for two-phase
  flow in porous media, SIAM Journal on Scientific Computing 41~(4) (2019)
  B746--B772.

\bibitem{Ren2016IJHMT}
Q.~Ren, C.~L. Chan, Numerical study of double-diffusive convection in a
  vertical cavity with soret and dufour effects by lattice {Boltzmann} method
  on {GPU}, International Journal of Heat and Mass Transfer 93 (2016) 538--553.

\bibitem{Huber2010JCP}
C.~Huber, B.~Chopard, M.~Manga, A lattice {Boltzmann} model for coupled
  diffusion, Journal of Computational Physics 229~(20) (2010) 7956--7976.

\bibitem{Chai2019PRE}
Z.~Chai, X.~Guo, L.~Wang, B.~Shi, Maxwell-stefan-theory-based lattice
  {Boltzmann} model for diffusion in multicomponent mixtures, Physical Review E
  99 (2019) 023312.

\bibitem{He2004CP}
N.~He, N.~Wang, B.~Shi, Z.~Guo, A unified incompressible lattice {BGK} model
  and its application to three-dimensional lid-driven cavity flow, Chinese
  Physics 13~(01) (2004) 40--46.

\bibitem{Holdych2004JCP}
D.~J. Holdych, D.~R. Noble, J.~G. Georgiadis, R.~O. Buckius, Truncation error
  analysis of lattice {Boltzmann} methods, Journal of Computational Physics
  193~(2) (2004) 595--619.

\bibitem{Wagner2006PRE}
A.~J. Wagner, Thermodynamic consistency of liquid-gas lattice {Boltzmann}
  simulations, Physical Review E 74 (2006) 056703.

\bibitem{Kaehler2013CCP}
G.~Kaehler, A.~J. Wagner, Derivation of hydrodynamics for multi-relaxation time
  lattice {Boltzmann} using the moment approach, Communications in
  Computational Physics 13~(3) (2013) 614–628.

\bibitem{Chai2020PRE}
Z.~Chai, B.~Shi, Multiple-relaxation-time lattice {Boltzmann} method for the
  navier-stokes and nonlinear convection-diffusion equations: Modeling,
  analysis and elements, Physical Review E 102 (2020) 023306.

\bibitem{Chai2016JSC}
Z.~Chai, B.~Shi, Z.~Guo, A multiple-relaxation-time lattice {Boltzmann} model
  for general nonlinear anisotropic convection–diffusion equations, Journal
  of Scientific Computing 69 (2016) 355--390.

\bibitem{Zhang2012PRE}
T.~Zhang, B.~Shi, Z.~Guo, Z.~Chai, J.~Lu, General bounce-back scheme for
  concentration boundary condition in the lattice {Boltzmann} method, Physical
  Review E 85 (2012) 016701.

\bibitem{Ladd1994JFM1}
A.~J.~C. Ladd, Numerical simulations of particulate suspensions via a
  discretized {Boltzmann} equation. {Part} 1. {Theoretical} foundation, Journal
  of Fluid Mechanics 271 (1994) 285--309.

\bibitem{Ladd1994JFM2}
A.~J.~C. Ladd, Numerical simulations of particulate suspensions via a
  discretized {Boltzmann} equation. {Part} 2. {Numerical} results, Journal of
  Fluid Mechanics 271 (1994) 311--339.

\bibitem{Xu2019IJTS}
H.~Yu, Z.~Luo, Q.~Lou, S.~Zhang, J.~Wang, Lattice {Boltzmann} simulations of
  the double-diffusive natural convection and oscillation characteristics in an
  enclosure with {Soret} and {Dufour} effects, International Journal of Thermal
  Sciences 136 (2019) 159--171.

\end{thebibliography}




\end{document}